# Spatial structure and composition of polysaccharide-protein complexes from Small Angle Neutron Scattering

*I. Schmidt[1], F. Cousin[2], C. Huchon[2], F. Boué[2], M. A.V. Axelos[1]*

[1] UR1268 Biopolymères Interactions Assemblages, INRA, F-44300 Nantes France

[2] Laboratoire Léon Brillouin, CNRS-CEA IRAMIS UMR12, CE Saclay, F- 91191 Gif-sur-Yvette Cedex, France.

*Abstract:*

We use Small Angle Neutron Scattering (SANS), with an original analysis method, to obtain both the characteristic sizes and the inner composition of lysozyme-pectin complexes depending on the charge density. Lysozyme is a globular protein and pectin a natural anionic semiflexible polysaccharide with a degree of methylation (DM) 0, 43 and 74. For our experimental conditions (buffer ionic strength $I = 2.5 \cdot 10^{-2}$ mol/L and pH between 3 and 7), the electrostatic charge of lysozyme is always positive (from 8 to 17 depending on pH). The pectin charge per elementary chain segment is negative and can be varied from almost zero to one through the change of DM and pH. The weight molar ratio of lysozyme on pectin monomers is kept constant. The ratio of negative charge content per volume to positive charge content per volume, -/+, is varied between 10 and 0.007.

On a local scale, for all charged pectins, a correlation peak appears at 0.2 Å$^{-1}$ due to proteins clustering inside the complexes. On a large scale, the complexes appear as formed of spherical globules with a well defined radius of 10 to 50 nm, containing a few thousands proteins. The volume fraction Φ of organic matter within the globules derived from SANS absolute cross-sections is around 0.1. The protein stacking, which occurs inside the globules, is enhanced when pectin is more charged, due to pH or DM. The linear charge density of the pectin determines the size of the globules for pectin chains of comparable molecular weights whether it is controlled by the pH or the DM. The radius of the globules varies between 10 nm and 50 nm. In conclusion the structure is driven by electrostatic interactions and not by hydrophobic interactions. The molecular weight also has a large influence on the structure of the complexes since long chains tend to form larger globules. This maybe one reason why

DM and pH are not completely equivalent in our system since DM 0 has a short mass, but this may not be the only one. For very low pectin charge (-/+ = 0.07), globules do not appear and the scattering signals a gel-like structure. We did not observe any beads-on-a-string structure.

*1. Introduction:*

The complexation or coacervation of polyelectrolytes (PEL) and proteins is of major interest in the field of biomacromolecules, because they are involved in biological processes such as interactions in plants or animals, but also because they are encountered in many industrial applications. Of particular importance are polysaccharides and proteins of opposite charges. They have been studied in detail over the last years since the pioneer works of Bungenberg de Jong [1,2] and Morawetz and Hughes [3], as is illustrated by the increasing number of recent reviews dealing with the subject [4,5,6,7,8]. Applications are the stabilization of emulsions [9], food industry, pharmacology, drug release, biosensors, micro-encapsulation and the release of proteins, polyelectrolyte multilayers or protein separation [see ref 4,5,6,7,8 and references therein]. In this context it is important to control the macroscopic properties of the mixtures (viscosity, colloidal stability…), and other parameters, all depending on the structure; this requires getting a fine description of the mechanisms of complexation. This is not trivial because many parameters can influence the interactions and thereby the complexation: pH, ionic strength, concentration, charge density of the polyelectrolyte, charge distribution on the protein, rigidity of the chain, protein to polysaccharide ratio, and relative hydrophobicity of the molecules.

In spite of this complexity, it is now recognized that electrostatic interactions are predominant for the complexation of these oppositely charged species in several model systems. In particular, it has been shown experimentally that aggregation is at its maximum when the stoichiometry of the charges is equal to 1 which corresponds to neutralization of the complexes [10, 11]. This leads to another step which also depends on the strength of the interactions: either liquid-liquid phase separation (coacervation) where one phase is enriched with soluble complexes, or liquid-solid phase separation (precipitation). Precipitation occurs for stronger interactions, i.e. for high charge densities, pH far from the isoelectric point pI of the protein or low ionic strength. When proteins are involved instead of homogeneously charged spheres, the presence of patches of opposite charges on the surface of the protein enables electrostatic complexation with polyelectrolyte of the same global charge. Indeed it has been shown that complexation can occur at a critical pH higher than the pI of the protein [12, 13]. Another key factor for the complexation in such charged systems like PEL and proteins is the gain of entropy associated with the release of the counterions (c.i.) condensed on the charged chains (Manning condensation) or spheres. Calorimetric experiments have indeed shown that complexation is mostly entropy driven [6, 10, 14]. Recently, in some systems globules were found as the elementary object in the complex, and the first direct structural measurement of the counterion location on these proteins/PEL globules has demonstrated that this gain of entropy is associated with the total release of c.i. from the core of the globules, which is neutral (inner charge stoichiometry) [15]. But it has also been demonstrated experimentally recently that complex coacervation could be enthalpy driven in some other systems [16, 17, 18]. Therefore it seems that the charge density of the polyelectrolyte is one of the main parameter in the complexation: highly charged PEL induce a major entropic contribution due to the release of condensed counterions or water molecules, whereas for weakly charged PEL one has a major endothermic enthalpic contribution.

The importance of the stoechiometric ratio, which we introduce below as [-]/[+], has been also considered (being noted $x = [+]/[-]$)by analytical theories, , in a reductionist approach which considers mixtures of large macroions with smaller Z- ions of opposite charge, both being possibly either spheres or chains[43, 44]. The case of large chains with spheres [44], is closest to the case reported here our work. In these systems, authors predict a "condensation", that is the formation of globules, in a central region of all diagrams, where $x$ is of the order of 1. In the case of polycation and polyanion chains [43], there can be coexistence between globules containing either type of chains, and free chains in excess, or an intracomplex "disproportionation" by formation of tadpoles with a spherical part and a string-like part. In the case of spheres plus chains (which can be flexible or semiflexible), two types of complexes are predicted: a partial aggregation as a bead-on-a-string type (called by authors "artificial chromatine"), and strong aggregation (condensation) in globules. The "strong aggregation" zone in the diagram is centered on the value x = 1, and surrounded by "partial aggregation" zones.

The role of electrostatic interactions in chains-spheres mixtures,  and their interplay with parameters such as chain stiffness was studied in parallel by computer simulations, at first in simulations involving only one protein and one polyelectrolyte chain [19, 20], and more recently on systems dealing with several proteins in the presence of an oppositely charged PEL [21, 22]. In ref [22], the heterogeneous charge density of proteins in the presence of weak PEL has also been taken into account whereby different PEL lengths and ionic strengths were employed. Large PEL-Protein clusters are formed by simulation for systems with the strongest electrostatic interactions at the stoichiometric charge ratio. With PEL in excess, re-dissolution occurs, with proteins still attached to PEL. Smaller clusters are formed by simulation with shorter PEL chains and for increased ionic strength. Complexation is promoted by decreasing the ionic strength. All these numerical studies underline the importance of the PEL stiffness: complexation is also promoted by making the PEL more flexible. These main trends have been checked experimentally by Kayitmazer *et al*. [23] that compare the binding of chitosan and PDADMAC, two PEL with same charge density but different persistence lengths, to oppositely charges micelles, dendrimers and proteins. However the authors point out that the relevant parameter for binding is not necessarily the intrinsic persistence length of the chain $L_p$, but the flexibility of the chain on the colloid length scale.

Concerning the influence of the ionic strength, comparison between simulation and experiments is still controversial. While some experimental studies show that the determining scale is the Debye length $\kappa^{-1}$ [7], in accordance with simulations, another experiment has shown recently that a decrease of $\kappa^{-1}$ (i.e. an increase of electrostatic screening), can increase the size of protein/PEL globules when they are present in the complexes [24]. This comes from the fact that there are two different scales for two different interactions: (i) the local scale of the direct interaction between protein and PEL, which is well described in the simulations, and (ii) the scale of the interactions between primary globules (once they are formed), which is not considered in simulations. At local scale, the asymmetry of electrostatics around strongly patched proteins play also a major role [25]. There is then an "optimal ionic strength" for binding, usually lying between 10mM and 30mM, when the Debye length is of the order of the protein radius because both electrostatic global repulsions between proteins and attractions between oppositely charged patches can occur.

Interactions other than electrostatics, though the latter are the most important for complexation, have to be considered. Examples are hydrogen bonding [4], e.g. for PEO - or polysaccharide - based backbones, and hydrophobic interactions if the backbone of the PEL is partially hydrophobic, e.g. due to alkyl chains [4, 26]. In the latter case, interactions with hydrophobic patches on the protein can contribute to formation of a complex [10, 27], but also to protein unfolding [28]. The "optimal ionic strength" [25] quoted above can also be tuned by hydrophobic interactions which act on the short range.

While the understanding of the mechanisms of complexation of PEL with proteins has substantially progressed over the last years, there remain some open questions for a perfect tuning of the properties. In particular kinetics aspects have to be considered because systems are often out of equilibrium. While many studies were done at thermodynamic equilibrium [16], kinetics is now beginning to be considered. The nucleation and growth mechanism is accepted as a general coacervation/complexation mechanism for acidification induced complexes between biological molecules [29, 30]. The evolution of the structure on the largest scale obviously depends on the physico-chemical parameters such as the charge density of the components: for example, for highly charged PEL, like PSS, the globules, which are the objects formed primarily, attract to finally precipitate whereas for lower charged PEL like pectins, coacervates are formed [6, 30]. It is likely that the evolution of the kinetics of the complexes during the evolution towards their final structure is mainly dependent on the local structure on the scale of the primary objects (10Å – 500Å, typically the one probed by simulations). A more refined structural description on these scales will bridge the gap between the simulations and the macroscopic experiments.

In this framework Small Angle Neutron Scattering (SANS) recently has shown to be a powerful technique for the determination of the structure of complexes: it is the method of choice to analyze the different relevant length scales of the system and the local density fluctuations, to be compared with other experimental techniques such as rheology, pH titration, light scattering or turbidimetry which were mainly used before. For example, two kinds of microstructures were proposed for β-lactoglobulin/pectin coacervates [32]: the first one involves protein molecules as distributed separately in the coacervate network, whereas the second one encloses larger protein domains in the entangled pectin chains. It has also been reported that agar-gelatin complex coacervates form physically crosslinked networks of polymer-rich zones separated by polymer-poor regions and a correlation length around 12Å [33]. SANS have also been combined with SALS to study the structure of BSA-chitosan coacervates [34], which appear formed of protein-PEL aggregates. These aggregates are largely solvated (they contain around 84% of water) which points out the role of PEL chain stiffness on compactness of protein-PEL aggregates since much denser aggregates (~ 75% of water) have been obtained in coacervates of the same protein with PDAMAC [35], a less rigid PEL (the intrinsic persistence length is ~ 25 Å and ~ 60 Å for chitosan).

When polymer could be deuterated, in the case of lysozyme – PSSNa complexes [11, 15, 24, 28, 31], the systematic use of contrast matching gave separately the scattering of the different species within the complexes: lysozyme, all PSS chains, isolated chains and even c.i.. Two main structures were determined [28] depending on the level of interpenetration of the chains [36]: (i) a gel where the initial transient network of PEL is crosslinked by proteins in the semi-dilute regime and (ii) dense spherical globules of a few hundreds of Å in size, containing both protein and PEL, and arranged at larger scale as a fractal network linked by PEL chains in the dilute regime. In this case the inner composition in the globules could be evaluated separately for the two species.

The aim of the present paper is to apply the technique of SANS to complexes made of a lysozyme and pectin, a natural polycarboxylate. Pectin is a polysaccharide which is always present in plant cell walls and which is commonly used as a gelling agent in the food industry. Lysozyme is a model protein to model interactions between polysaccharides and other basic proteins, such as plant proteins. Plant proteins are very often encountered in combination with pectin during fractionation of agricultural product or in food formulation. We will show that SANS can be used to evaluate quantitatively the concentrations of the species inside the complexes despite the fact that pectin cannot be deuterated. We will especially test the influence of the charge densities of the species, and in particular of the charge ratio in the mixture $[-]/[+]_{intro}$ and the influence of the pectin molecular weight on the local structure of the pectin/lysozyme complexes.

*2 Materials and methods*

**2.1 Materials**

Hen egg lysozyme is an ellipsoidal globular protein with 128 residues [37] and a molecular weight of 14298 g/mol. Its dimensions from crystallographic data are 30 x 30 x 45 Å, and in solution its form factor gives a radius of gyration $R_g \sim 17$ Å. Its isoelectric point is at pH 10.7 and it is positively charged at low pH. It carries 28 ionisable groups, which are ionized into 10 negative charges at high pH, and into 18 positive charges at low pH. The net charge of the protein varies between +17 and +8 depending on the pH conditions as given in Table 1 of the Supporting Information. The ionic strength of the co-ions is taken to be equal to the molarity times the net charge. Our sample has been purchased from Sigma (L-6786, lot n°111H7010) and used without further purification.

Pectin consists mainly of a semi-flexible backbone of α-(1→4) linked -D-galacturonic acid and its methyl ester. The charge density of the polymer is proportional to the number of non methylated monomers per 100 monomers units, i. e. (1 – DM)/100 where DM is the degree of methylation (DM). It also depends on the pH as the dissociation coefficient of the carboxyl groups is pH-dependent (they are fully dissociated at pH 7.4). The number of charges (COO-) per gram depends on the pH conditions and the pectin characteristics, and has been measured directly by pH titration in all cases. It is presented in Table 1 of the Supporting Information (S.I).

In this paper pectin has three different DMs. The notation DMnn will be used to differentiate the pectin samples. DM43 and DM74 have been graciously provided by Copenhagen Pectin; DM0 has been purchased from SIGMA. The commercial samples were purified in order to eliminate potential traces of residual parietal proteins, salts or cations like $Ca^{2+}$. Pectins were put in acidic form in ethanol acidified by HCl: 5g of pectin powder were dissolved in 100 mL of a solvent mixture composed by 70% (v/v) of ethanol and 5% (v/v) of HCl. After stirring during 1 H at ambient temperature, the powder was recovered on a fritted glass (porosity n°4) and successively washed with solvent mixtures of ethanol (at 70%, then at 95%) and finally with absolute ethanol. The samples were dried during 48h under the hood and then kept at 4°C.

The molecular mass has been determined by a combination of HPSEC-MALLS measurements in $NaNO_3$ 0.05M (High performance Size Exclusion Chromatography – Multi Angle Laser Light Scattering) [39] and by measurements of the intrinsic viscosity, a safer method since it is less sensitive to aggregates. Pectins DM74, and DM43 have almost the same value of the intrinsic viscosity [η] in 0.1 M NaCl at 20°C: 0.313 L.g$^{-1}$, 0.336 L.g$^{-1}$ respectively; they have an average molecular weight $M_w$ of about 90000 g/mol with a polydispersity index of 2 (~ 460 monomers per chain). The intrinsic viscosity of DM0 is considerably lower: 0.071 L.g$^{-1}$, which corresponds to a molecular weight for DM0 around 17 000 g/mol (~ 100 monomers). This is attributed to backbone degradation during de-methylation. The values of the intrinsic viscosity allow to evaluate the overlapping concentration c* = 1/ [η] = 14 g/L for DM0 and 3g/l for both DM74 and DM43. Persistence lengths measured formerly for similar pectins [38] have been reevaluated for our solutions (see section Results).

All experiments presented hereafter are performed on mixtures realized with $C_{pectin}$ = 5.5 g/L and $C_{lysosyme}$ = 5 g/L. For DM74 and DM43 the pectin concentration is slightly higher than the overlapping concentration c*, separating the dilute from the semidilute regime. For DM0 chains are in the dilute regime. At these concentrations, the volume fraction of pectin and lysozyme are very close: $\Phi_{pectin}$ = 0.00314 ($\rho_{pectin}$ = 1.75 g/ cm$^3$) and $\Phi_{lysozyme}$ = 0.00305 as determined by SANS (see Supporting information). In the following is studied the influence on the interactions between lysozyme and pectin of: (i) the pH, using three pH values close to 3, 4 and 7 (the exact values depend on the sample) and (ii) the methylation ratio using DM74, DM43 and DM0.

Prior to mixing, solutions of pure pectin at 11 g/L and pure lysozyme at 10 g/L are obtained by dissolution in the same buffer, which is a mixture of an aqueous solution of $KH_2PO_4$ (4.36 $10^{-3}$ mol/L) and of an aqueous solution of $Na_2HPO_4 2H_2O$ (6.73 $10^{-3}$ mol/L). For SANS experiments the buffers have been obtained by dissolution of salts in a pure solution of $D_2O$ (purchased from Eurisotop, France). The global ionic strength of the buffer is 2.5 $10^{-2}$ mol/L, with the possibility to adjust the pH to the required value by adding NaOH or HCl. Lysozyme solutions are gently stirred for homogenization for 10 minutes. Pectin solutions are stirred for 24 hours to complete solubilisation. Two equal volumes of each of the solutions are then mixed and homogenized with a vortex for a few seconds and then left at rest at 4°C for at least two days. Table 1 gives, for each mixture studied, the number of negative charges, the ratio $[-]/[+]_{intro}$ and the total ionic strength ($I=I_{buffer} + I_{lyso} + I_{pect}$). Two samples have been measured at two different times, DM0 and DM74 at pH 7, which will be useful to test reproducibility. The ionic strength from lysozyme has been estimated as the molarity times the number of charges for the pH 3 (Figure S. I. 1 in S.I.). Its order of magnitude does not vary strongly with pH. Only at higher pH the pectin contribution is more important and can increase the ionic strength by a factor 2 compared to the pH of the buffer.

One can note here that since we work at a single chain concentration, the parameter $[-]/[+]_{intro}$ is also proportional to the ratio "average linear density of charge on the chain over the average charge of the protein".

When solutions of lysozyme and pectins are mixed, a turbid mixture is obtained instantaneously that strongly scatters light, except for DM74 at pH 3. Two days later, some turbid samples show separation into either a turbid plus a clear phase, or into two turbid phases of different turbidity, one of which is a precipitate.

| Pectin | pH | $[-]/[+]_{intro}$ | I (M) | Visual aspect |
|---|---|---|---|---|
| DM0 | 7.2 | 10 | 0.056 | Turbid |
| | 3.75 | 3.36 | 0.043 | Turbid |
| | 2.9 | 0.47 | 0.034 | Turbid; clear supernatant $\Phi_{clear}$ ~ 0.04 |
| DM43 | 7.2 | 3.6 | 0.036 | Turbid |
| | 4.1 | 1.37 | 0.036 | Turbid[*] |
| | 2.8 | 0.13 | 0.032 | Turbid; clear supernatant $\Phi_{clear} \sim 0.04$ |
| DM74 | 7.1 | 1.6 | 0.033 | Turbid |
| | 4.1 | 0.65 | 0.03 | Turbid; Turbid precipitate $\Phi_{precipitate} \sim 0.04$ |
| | 3 | 0.07 | 0.0315 | Clear |

* Phase separation possible on a 24h time scale (see text).

Table 1: Samples studied by SANS for **constant** $C_{pectin}$ = 5.5 g/L and $C_{lysosyme}$ = 5 g/L at different pH: charge of species, charge ratio (which is also proportional to the average linear density of charge on the chain over the average charge of the protein since we work at quasi-constant chain concentration), ionic strength (I(M) = $I_{buffer}$ + $I_{lyso}$ + $I_{pect}$), and visual aspect. For DM0 and DM74 at pH ~ 7, two different samples have been measured at two different times.

Proteins in a free or a complexed state have been measured by colorimetric and UV detection respectively after separation using ultra-filtration membrane (Amicon Ultra-4, 100 000 MWCO, Millipore, USA) for different concentrations. These experiments [40] were carried out at pH 7 for three DMs, 0, 43 and 74. They show that more than 98% of the proteins are in the complexes of the turbid mixtures or in precipitates when the latter are observed.

### 2.2 SANS measurements

SANS measurements were performed on PACE and PAXY spectrometer (Orphée reactor, LLB, CEA Saclay, France). Three different configurations were used with two different neutron wavelengths and two sample-detector distances ($\lambda$ = 13 Å, D = 4.7 m; $\lambda$ = 6 Å, D = 4.7 m; $\lambda$ = 6 Å, D = 1 m) to cover a q-range of over two decades: 0.00245 Å$^{-1}$ < q < 0.366 Å$^{-1}$ (for a first series measurement, the one shown below for pH 7"), and 0.0035 Å$^{-1}$ < q < 0.338 Å$^{-1}$ (including all other pHs and a second series at pH 7 used a test for reproducibility). All measurements presented in the main text were performed in D$_2$O. Table

S.I 2 in the Supporting Information gives the scattering length density ρ (cm$^{-2}$) of all the species under study in D$_2$O.

Standard corrections [41] for sample volume, neutron beam transmission, empty cell signal and detector efficiency have been applied to get the scattered intensities in absolute scale for the samples with complexes, I$_{sample}$(q), and for the pure deuterated solvent I$_{solvent}$(q). The scattering from the complexes I$_{complexes}$(q) is obtained by subtracting the solvent signal:

$$I_{sample}(q) = \Phi \cdot I_{complexes}(q) + (1-\Phi) \cdot I_{solvent}(q) \qquad (1)$$

where Φ is the volume fraction occupied by the objects of the complexes.

All SANS measurements have been performed two days after samples preparation.

One has to take into account the possibility of precipitation or slow decantation inside the cell (see Table 1), which removes part of the complexes off the beam. The samples were gently shaken before putting them into in the sample changer. No precipitation was noticed neither before the recording of the spectrum nor when removing samples between 12 H and 24 H later, except for DM43 at pH 4.1. But even in this case we are confident of the result, for the following reason: an important precipitation, by removing a fraction of the material off the beam would be noticeable in the large q range. There the signal of all lysozyme, complexed or free, has been measured such that the scattering is equal to the one of a solution of free soluble lysozyme at the same concentration.

*3 General features of the scattering*

**3.1 Individual components of the mixtures**

Pure solutions of pectin at 10 g/L and of lysozyme at 10 g/L have been measured by SANS to test the initial dispersion state of objects in the solution. The solvent is the same buffer in D$_2$O as for pectin/lysozyme mixture at pH 7. Figure 1 presents the scattering intensities obtained for lysozyme, and DM74 respectively.

In the large q domain the scattering from lysozyme is much larger than the one from pectin. This is due to the difference of the scattering law of objects in this Porod domain and to the higher neutronic contrast of lysozyme compared to pectin in D$_2$O (here $\Delta\rho_{lysozyme}^2 \sim 2\, \Delta\rho_{pectin}^2$). This has a strong influence on SANS experiments of mixtures as the signal at large q is always dominated by the signal of the lysozyme. The volume of the protein V$_p$ (14465 Å$^3$) obtained from scattering of lysozyme solutions is in good agreement with crystallographic measurements (See Supporting Information). At low q a slight upturn due to aggregates is observed for lysozyme, but is not large enough to compete with the scattering from complexes.

SANS measurements obtained for DM74 and DM43 show a classical behavior of polyelectrolyte chains in solution (here for clarity we present only the results for DM74 only: a transition between a $q^{-1.7}$ behaviour at low q (the excluded volume exponent is 1.7 for self avoiding chain behavior) and a $q^{-1}$ behaviour at larger q (rod-like behavior). Extracted from the cross-over abscissa $q_{c.o.}$ the typical values of the persistence lengths $L_p = 6/(\pi.q_{c.o.})$ in a 25mM buffer are of the order of 80 Å for DM74 at 10g/L and for DM43 at 4 g/L, slightly higher than what was obtained from earlier investigations at higher ionic strength [34].

**[Figure 1]**

**SANS spectra of individual components of the mixture: ■ lysozyme (10g/L); ▲ pectin DM74 (10g/L).**

### 3.2 Mixtures

Figure 2 present the scattering intensity obtained from the mixtures as a function of the pH, for each of the three DMs.

**[Figure 2]:**

**SANS spectra for lysozyme-pectin complexes for each of the three pectins (DM0, DM43 and DM74) as a function of pH: (a) pH ~ 7, (b) pH ~ 4, (c) pH ~ 3. Each set of spectra is compared to the spectrum of free lysozyme at the same concentration as in the mixture (5g/L).**

Most spectra present the same features.

- at large q, I(q) decreases as $q^{-4}$. Such a behavior is explained by a "Porod scattering" associated with a sharp interface between compact objects and the solvent. Here, this decrease is the same as for pure dilute lysozyme, which indeed is a compact globular protein. At this large q scale, i.e. short distances, we expect to see the individual contribution of the two components. We have seen in Figure 1 that at q > 0.01 Å$^{-1}$ the individual scattering of lysozyme is much larger than for pectin. Thus in the mixtures the scattering reflects mostly the lysozyme on this scale, whether it is free or inside complexes.

- around 0.2 Å$^{-1}$, all samples except DM74 at low pH display a maximum, which is more or less marked depending on the sample. This maximum is still due to lysozyme, and signals a correlation size. Its abscissa is 0.2 Å$^{-1}$ in most of the cases, which corresponds to a mean distance $2\pi/0.2 \sim 30$ Å equal to the distance between two proteins in close contact (via their smallest axis [28]), which we call "stacked". For DM43 at pH ~ 3 (there is also a minute effect at pH = 4.1), the correlation peak is shifted toward lower q, 0.16 Å$^{-1}$, corresponding to 40 Å. In this case, the stacking is looser, and leads to an average between different configurations of protein-protein sticking.

- going to lower q, the intensity increases again. It is much larger than the scattering of the individual lysozyme, which shows a plateau in this range, and also larger than the pectin scattering, which scatters as $q^{-1.7}$. Here the slope is again -4 (except for DM74 at pH3, which makes the scattering increase not as fast when decreasing q, with an apparent slope 2.1). This second $q^{-4}$ decay (therefore also a "Porod scattering") is due to the surface scattering of objects that are larger than the individual components. These objects are large and compact. Their scattering is very close to the one of primary objects observed in simulation and in scattering from other mixtures. It has been seen in particular for PSS, and correlated with Electronic Microscopy after cryofracture showing quasi-spherical globules which can contain both lysozyme and pectin. Hence we will call these objects "primary globules".

We can note at this stage that we see no sign of elongated structure (which would result in a $q^{-1}$ or at least to a slower decrease than $q^{-4}$ at the lower q border of the $q^{-4}$ range), as could show "tadpoles" [43], or, as predicted for systems of chains plus spheres, a bead-on-a-string (as artificial chromatine in [44]) type of structure.

- finally, at even smaller q, there is a change in the scattering curve on a log-log scale: it does not show a power law -4 any longer, but a downturn shoulder. The downturn can then be due to the finite size of the primary complexes. For DM 0 at pH 3.8 and 2.9, a new, lower, slope appears (~ 2.5). In this case it maybe due to a crossover to a new regime of spatial arrangement of the primary globules. The next section gives a detailed qualitative comparison, and shows that a quantitative analysis of the composition of the primary globules is possible from the Porod scattering.

Before proceeding with the details, two important features of the scattering curves should be pointed out. First, at large q all curves of Figure 2 overlap with the free lysozyme data. In this range one probes the form factor of the objects; and the scattered intensity is almost constant in first approximation. Since the pectin scattering is negligible in this q range, the scattered intensity comes from the proteins which are mainly surrounded by water (see end of paragraph 4.1), and thus have a contrast that is the same whether they are in globules or not. This means that the amount of specific area of the proteins is constant from one sample to another. As the same amount of proteins has been introduced in all samples, this proves that the lysozyme concentration probed during the experiment is the same as the one introduced in the sample (no precipitation).

Second, Figure S. I. 2 of the Supporting Information shows some experiments where we matched the scattering of either lysozyme in a 43% $D_2O$/57%$H_2O$ mixture, or pectin in a 60% $D_2O$/40%$H_2O$ mixture. The contrast in both cases is the one between lysozyme and pectin, $\Delta\rho^2 = (\rho_{lyso} - \rho_{pect})^2 \sim (1.5 \; 10^{10} \; cm^{-2})^2$, seven times lower than the contrast between lysozyme and heavy water ($\sim 4 \; 10^{10} \; cm^{-2})^2$. The contrast is too weak to allow a measurement with correct statistics in a reasonable time for $q > 0.02$ Å$^{-1}$. One nevertheless sees that the two signals overlap at low q and the curve has the same shape as the one between lysozyme and heavy water. Since pectin and lysozyme have approximately the same volume fraction, it shows that both species are located inside the same compact objects - the globules- to an equal degree.

## *4 Analysis of the scattering*

### 4.1. Qualitative descriptions

First is described just below the effect of DM (which influences both the linear charge density and the hydrophobicity of the chain), at constant pH, as seen in Figure 2.

At pH ~ 7 (Figure 2.a), the intensity of the scattering in the $q^{-4}$ range is lowered when the ratio [-]/[+]$_{intro}$ is increased, which suggests, according to the Porod law, that the specific area decreases assuming that the contrast between globule and solvent does not change (this will be quantitatively discussed below). This suggests that the globule size increases with [-]/[+]$_{intro}$. This could be checked at lower q, by the variation of the q abscissa at which the downturn occurs, but we cannot see whether it occurs at different q for the three samples in this representation. Simultaneously, at large q, when [-]/[+]$_{intro}$ is increased, the correlation peak at 0.2 Å$^{-1}$ is more pronounced, which suggests a more important protein stacking.

At pH ~ 4 (Figure 2.b), the large q behavior varies in the same way: the correlation peak is slightly more pronounced for DM 0, which is more charged. But at low q the order for the size of DM 0 and DM 43 is inverted compared to pH 7: DM 0 displays a $q^{-4}$ behaviour with a **higher** intensity than DM 43 at intermediate q, suggesting a smaller size for higher $[-]/[+]_{intro}$ (again assuming equal contrast with solvent for the globules of both DMs). Here this is confirmed clearly, (at opposite with pH 7), by the change of slope of the curve at small q, which arises sooner for DM 0, around 0.01 Å$^{-1}$ instead of 0.007 Å$^{-1}$ for DM43. This inversion in the order can be correlated with the low molecular weight of DM 0, as discussed below.

At pH ~ 3 (Figure 2.c) pectins are weakly charged, $[-]/[+]_{intro}$ becomes lower than 1 for all DMs. But DM0 ($[-]/[+]_{intro}$ 0.47) and DM43 ($[-]/[+]_{intro}$ 0.13) compare in a similar way as for pH 4: at intermediate q the scattering of DM0 is again higher, meaning smaller globules at equal contrast. At low q, this is confirmed: DM0 leaves the Porod regime and enters a regime of q smaller than the inverse globule size at a larger crossover value (q = 0.02 Å$^{-1}$)than DM43 (q = 0.01 Å$^{-1}$)). So that the DM0 scattering passes below the DM43 one. At large q, the correlation peak depends on the charge as observed for the other pHs: like at pH 4, it evidences for DM 43, a less perfect stacking of the proteins (peak shifted towards low q).

We were expecting, at pH 3, strong effects to appear, drastically, since the charge become very weak for some of the DMs. This is true only for DM 74 ($[-]/[+]_{intro}$ 0.07): by contrast to DM0 and DM43, striking changes are seen in the scattering: (i) at low q: the slope is much lower (-2.1); (ii) at large q there is no maximum, no correlation peak: the signal superimposes on the one of lysozyme in a dilute solution, i.e. there is no spatial organization of the protein on a local scale. In summary, when $[-]/[+]_{intro}$ is very low, but only in this case, the globule cannot form.

**[Figure 3]**

**SANS spectra for lysozyme-pectin complexes for the three pH values (pH ~3, pH ~4 and pH ~7) as function of DM, at large q : (a) DM 0, (b) DM 43, (c) DM 74. and at low q (d) for DM0. Each set of spectra is compared to the spectrum of free lysozyme at the same concentration as in the mixture (5g/L).**

We now follow the effect of the pH at constant DM. When observing DM0 as a function of pH (Figure 3.a), we see the same dependence on $[-]/[+]_{intro}$ as noted above: the lower the charge, the higher the signal in the $q^{-4}$ range, and the lower the size of the globules. We also show in Figure 3.d the variation of I(q) in the low q regime: this enables to see clearly the change in slope at q at 0.01 Å$^{-1}$ and 0.02 Å$^{-1}$ for pH 2.8 and 3.75, discussed above, which induces a crossing of the curves. The behavior for DM 43 is different from the one for DM 0, at low q as already seen above, but also at large q as seen now in Figure 3.b. The evolution with pH of DM 74 is given in Figure 3.c; it also gave us the opportunity to check the reproducibility by comparing two series of measurements at pH 7.1 and 7.5 that show that the superposition is very satisfactory, knowing that all preparation and measurement steps are distinct. This also true at low q (as can be seen more clearly in Fig. S.I.3.c of Supporting Information), and we find the same behavior all over the q range for two DM0 samples also prepared in two different sets of experiments (Fig. S.I.3.a).

**[Figure 4]**
**Comparison of SANS spectra for lysozyme-pectin complexes at constant charge ratio (a) $[-]/[+]_{intro}$ ~ 3.5 and (b) $[-]/[+]_{intro}$ ~ 1.5.**

Finally, we present comparisons of SANS spectra at close charge ratio $[-]/[+]_{intro}$ but different DM. Namely, we choose two samples for which the DM are different, but two different pH values are chosen such that the degree of ionization is larger for the lower DM, resulting in the same global charge of pectin for both DMs. This is possible for two couples of samples. First, DM 0 at pH 3.75 and DM 43 at pH 7.2 have the same $[-]/[+]_{intro}$ ~ 3.5. In Figure 4.a, we see that (i) the low q behaviors are different because DM0 gives smaller globules (remember DM0 has a lower mass), (ii) the correlation peak at 0.2 Å$^{-1}$ is slightly more marked for lower DM (higher charge) showing slightly stronger stacking. The same reinforcement of stacking at lower DM is observed at large q on Figure 4.b when we compare DM 43 at pH 4.1 and DM 74 at pH 7, which both correspond to $[-]/[+]_{intro}$ ~ 1.5 (in this case there is no important difference at low q; keep in mind that molar masses are similar for DM43 and DM74). Therefore at same global charge of the chain, even for identical mass, there is an additional influence of DM, apart from acting on the charge. Whether it is due to hydrophobicity or to the distribution of charge along the chain will be discussed below

Finally we can summarize our analysis of the scattering by a simple and naïve sketch in real space, as done in Figure 7 shown at the end of the paper.

### 4.2. Quantitative analysis

*4.2.1 Porod's plots: determination of apparent radii, specific areas, apparent contrast and inner composition.*

In this section we show that we can extract both the apparent size of the primary complexes formed by complexation and their inner composition (the protein, polymer and water contents) from the scattering data because the measured scattering intensity is in absolute units.

**[Figure 5]:**

**Two examples of $I(q)q^4/2\pi = f(q)$ representations (log-log) at constant DM and constant pH (a) DM 0 at all pH, lower q range; (b) all DMs at pH 7, full q range.**

We use plots of $q^4I(q)$ (in the low q range in Figure 5.a, over the whole q range in Figure 5.b) which allows to better visualize the low q Porod scattering. In this q range, the plot of $I(q).q^4/2\pi$ displays two features: a smooth maximum followed by a plateau. This is characteristic of compact objects of well defined size; here we use the sphere as a model of such objects.

The position of the $q^4I$ maximum $q_{max}$ is linked to the size of the globules. The maximum corresponds to the first maximum of the product $P_{comp}(q)q^4$, where $P_{comp}(q)$ is the form factor of a spherical globule of radius $R_{comp}$ :

$$P_{comp}(q) = \left(3\frac{(\sin qR_{comp} - qR_{comp}\cos qR_{comp})}{(qR_{comp})^3}\right)^2 \qquad (2)$$

Such maximum is obtained for $q_{max} \cdot R_{comp} = 4\pi/4.52$. Table 2 reports the values obtained for the different complexes $R_{comp}$. Here again we have no sign of elongated structure. The maximum is observed nicely on all samples except for DM74 at pH ~ 3 (not shown), in agreement with the fact that in this case we directly deduced from the scattering the absence of compact primary objects in the complexes (Figure 2.c, no $q^4$ Porod scattering).

| Pectin | pH | $[-]/[+]_{intro}$ | $R_{comp}$ (Å) | $Iq^4/2\pi$ $10^{23}cm^{-5}$ | $R_{comp}.Iq^4/2\pi$ | $\phi_{comp}$ | $\phi_{inner}$ |
|--------|----|-----|----|----|----|----|----|
|        |    |     |    |    |    |    |    |
|      | 7.2  | 10   | 485         | 3.7 | 1800 | 0.087 | 0.07  |
| DM0  | 3.75 | 3.36 | 185         | 15  | 2800 | 0.063 | 0.1   |
|      | 2.9  | 0.47 | 110         | 25  | 2750 | 0.056 | 0.11  |
|      | 7.2  | 3.6  | 425         | 6.6 | 2800 | 0.066 | 0.095 |
| DM43 | 4.1  | 1.37 | 325         | 5.2 | 1690 | 0.095 | 0.065 |
|      | 2.8  | 0.13 | 280         | 10  | 2526 | 0.066 | 0.093 |
| DM74 | 7.1  | 1.6  | 405         | 7.4 | 3000 | 0.058 | 0.11  |
|      | 3    | 0.07 | No globules |     |      |       |       |

**Table 2: Results of the Porod scattering analysis: radius of globules ($R_{comp}$) and inner volume fraction ($\phi_{inner}$) were obtained from the $I(q)q^4 / 2\pi = f(q)$ representation at low q. $q_{max}$ is the abscissa of the maximum of $I(q)q^4 = f(q)$ plot, $R_{comp} = 4\pi \cdot q_{max} / 4.52$ the radius of the globule, $I(q)q^4/2\pi$ is the value of the plateau at higher q than the maximum, $\phi_{comp}$ the volume fraction occupied by the globules, $\phi_{inner}$ the inner volume fraction of organic species (lysozyme and pectin) within the complexes, assuming they are all inside the globules (so that $\phi_{inner} = (\phi_{lys} + \phi_{pect})/\phi_{comp}$).**

The volume fraction of species (protein plus pectin) $\Phi_{inner}$ can be obtained from the scattering curves with an original trick based on the measurement in absolute values that is fully developed in Appendix. The idea is to obtain the contrast $\Delta\rho_{comp}^2$ between the solvent and the globules, from the expression of the absolute intensity in the Porod regime (Eq. A.1), which involves $\Delta\rho_{comp}^2$ and an average radius of the globules which is we identify to $R_{comp}$. It is based on the hypothesis that all species are inside the globules, which has been checked by titration for proteins (see above). This contrast $\Delta\rho_{comp}^2$ is sensitive to the fraction of solvent in the globules.

Values are reported in Table 2. We obtain values from 6.5% to 11% for the inner fraction of pectin or lysozyme, and thus from 89 to 93% for the inner fraction of $D_2O$. The compactness is thus essentially constant from one sample to another. It seems to increase slightly at low pH, as long as globules still exist.

If we look at $q^4I(q)$ vs. q in the whole q-range like in Figure 5.b, we observe two flat regions in two different q ranges. The low q one, as discussed above, gives us some of the values of size and composition listed in Table 2. The second one, at the highest qs, corresponds to the Porod law from each lysozyme, corresponding to the scattering from the inside of the globules. The value must yield the specific area of lysozyme within the complexes. In this q-range, the scattering is thus proportional to the contrast $\Delta\rho_{lyso\_complexes}^2$ between the lysozyme and its D$_2$O/pectin environment within the globule. Since we use always the same amount of proteins, the resulting change in the plateau $q^4.I(q)$ level, from one sample to another, would be due to a change of lysozyme environment. It is less than 15% for our full set of samples in the accuracy of the experiment. This suggests, since both lysozyme and pectin inner concentrations are low inside the aggregates, that lysozyme is mainly surrounded by water –and not by pectin- in all cases.

### 4.2.2 Aggregation number

From the size of the globules and the lysozyme volume fraction within the complexes ($\Phi_{lyso} \sim 1/2\ \Phi_{inner}$), we can deduce the volume occupied by lysozyme per globule, and thus the number of lysozyme per globule $N_{agg}$:

$$N_{agg} = (\Phi_{inner}/2) \cdot (4/3\ \pi\ R_{comp}^3)/V_{lyso} \qquad (3)$$

We will focus here only on values for pH 7, in order to give some example of the variation. We take $\Phi_{lyso} \sim 1/2\ \Phi_{inner} \sim 0.05$ and $V_{lyso}$ approximately equal to 20 000 Å$^3$. We find that globules involve more than a thousand lysozyme molecules. For DM0 at pH 7, the value $R_{comp} \sim 486$ Å leads to $N_{agg} \sim 1500$ lysozyme in a globule, and for DM43 and DM74 at pH 7, the values $R_{comp} \sim 425$ and 405 Å leads to $N_{agg}$ close to 1000.

### 4.2.3 Intensity at very low q : interactions between globules

As the globules are centrosymmetrical in average, the intensity scattered by the globules can be written [11, 41]:

$$I(q)\ (cm^{-1}) = \Phi V_{comp} \Delta\rho_{comp}^2 P_{comp}(q) S_{comp}(q) \qquad (4)$$

where $\Phi$ is the volume fraction, $\Delta\rho_{comp}^2$ is the contrast (cm$^{-4}$), $V_{comp}$ the complexes volume (cm$^3$), $P_{comp}(q)$ the globule form factor and $S_{comp}(q)$ the structure factor between globule.

The structure factor $S_{comp}(q)$ can be evaluated by dividing I(q) by $\Phi V_{comp} \Delta\rho_{comp}^2 P_{comp}(q)$ (the values of $\Phi V_{comp} \Delta\rho_{comp}^2$ are taken from Table 2, and $P_{comp}(q)$ is calculated according to equation 2 valid for spheres (but, in the Guinier regime of $P_{comp}(q)$, $qR_g <1$, this treatment does not depend on the globule shape).

Figure 6 shows an example, for the pH 7 case, where the globules are well defined in size. We do not show the data for DM0: due to the larger values of $R_g$ in this case, measurements at even lower q would be necessary to establish the result. For the samples shown, the structure factor S(q) is larger than 1 and increases when q tends to 0, indicating a high osmotic compressibility of the dispersion of complexes. There are **effective attractions** between the globules. They are larger for DM43, which is the most charged. This suggests a correlation with electrostatics in such attractions, though they work differently from the ones which build the globules. It is also possible that the fact that the globules are smaller for DM43 makes the structure factor appear larger in the same q range.

For lower sizes of globules, such as for DM0 at pH 3.8 and 2.9 ($R_{comp}$ = 110 Å and 185 Å), we have a larger range of q values below 1/R. 1/R is signaled by the onset of the change in slope. The scattering goes on increasing at low q, still close to a power law in q, but with lower exponent. Thus the interactions here also are attractive, but, in addition, the scattering is close to the one of aggregates of apparent fractal dimension $D_f \sim 2.2$ to 2.5 (yielding a scattering in $q^{-Df}$). A scattering of this type (exponent 2.5) was observed for PSS-lysozyme complexes by SANS and the branched shape confirmed by TEM after cryofracture [31].

**[Figure 6]**
**Structure factor S(q) for mixtures of lysozyme and pectin at very low q at pH 7 for degree of methylation 74 and 43.**

**[Figure 7]**
**Tentative sketch of globular pectin-lysozyme complexes and their scattering in log-log plot (the $q^{-Dapp}$ part at low q is seen only in DM0 samples).**

*5. Discussion.*

## 5. 1 Influence of degree of methylation, pH, $[-]/[+]_{intro}$ and molecular weight on the structure of the pectin/lysozyme complexes

We will now discuss the scattering data as a function of the different parameters (DM, pH, molecular weight) using mostly the degree of stacking of the proteins, observed from the correlation peak at large q and the characteristics of the globules (size radius and compactness) obtained at low q. The effects of the different parameters are cross-correlated: remember that both DM and pH control the linear charge density of pectin, even though the distribution of charges along the chain is different in the two cases. Also since we work at quasi-constant chain concentration, the effect of the charge ratio and the effect of charge linear density are not disconnected. Thus we will try to focus first on the most pertinent correlations, and to finish with the apparently contradictory effects.

**Let us first focus on the correlation peak around 0.2 Å$^{-1}$.** There is an obvious correlation in the aspect of the maximum. However, the "landscape" appears different when looking at it from different angles. First, in Figure 2 it is easy to find the correlation with the degree of methylation DM at given pH. Figure 2.c at pH 3 shows the only case with no protein stacking: at large q the curve superimposes the free protein scattering data. This is because for the pectin is very weakly charged (DM 74 at pH 3, $[-]/[+]_{intro}$ = 0.07). In the same Figure, one sees that for any higher $[-]/[+]_{intro}$ (lower DM, DM 43 and DM 0) the peak appears immediately and gets more and more pronounced. In Figure 2.b, at pH ~ 4, this correlation in DM fades out a bit, but in Fig. 2a at even higher pH, 7, the effect of DM is still the same. Note that at this pH the $[-]/[+]_{intro}$ is high for all DM and becomes very high for DM0 ($[-]/[+]_{intro}$ =10). Thus at first sight we could conclude that the improvement of the peak is just an effect of the linear charge density due to decrease of methylation. However, in Figure 3, we see that another way of increasing the linear charge density, by tuning the pH instead of the DM, can have the opposite effect: for DM = 0 (Fig.3.a) and DM = 43 (Fig. 3.b), a high pH makes the peak less well defined, as can be seen clearly in Figures 3.a and 3.b. Since this is not observed for DM 74 at pH 7, which is less charged, we propose that it is due to too strong electrostatic interactions which produce an immediate but imperfect protein stacking by preventing complete rearrangements.

In Figure 4, the comparison of the SANS spectra at constant $[-]/[+]_{intro}$ (~ 3.5 in Fig.4.a for DM0 at pH 3.75 and DM43 at pH 7.2, and ~ 1.5 in Fig. 4.b for DM43 pH 3.75, and 1.6 for DM74 pH 7.1) yields an important information. In both cases, we see that the peak increases when DM decreases. That would be the opposite if putting in contact proteins was mostly due to hydrophobic attractions (if we assume in first approximation that electrostatics forces are constant at constant $[-]/[+]_{intro}$). So hydrophobicity does not win over electrostatics. One has nevertheless to find an explanation of the observed effect, namely that a sample of given constant $[-]/[+]_{intro}$ shows a looser stacking when its DM is large: it could be that the existence of uncharged sequences disorganizes the stacking, probably because it restricts the accessibility of the charged units.

**Let us now discuss the scattering in the low q domain** to test the influence of pH, DM and molecular weight on the globules shape, size, and compactness. All Figures 2, 3 and 4 show that the $q^{-4}$ behavior of the scattering at intermediate q, signature of the appearance of globules (as soon as $[-]/[+]_{intro} > 0.13$), is always associated with the presence of a peak at large q, showing that the structure of globules is always the same. Reciprocally, Figure 2.c shows that we do not observe globules in the only sample without peak, DM74 at pH 3 where the charge density is very weak ($[-]/[+]_{intro} = 0.07$). From the apparent exponent of q in the variation of I(q), we infer that the structure in this last case is the same as the one of an inhomogeneous gel observed in PSS/lysozyme systems [28]. In our case, the inhomogeneities remain probably limited to smaller scales (not accessible by SANS), as signaled by the fact that the samples remain optically clear.

We can follow the effect of pH at given DM, either using Figures 2.a, b, c (i.e. taking the same symbol shape through the three Figures), or using Figure S.I.3. But this is more simple using Table 2: we see that the increase of pH, i.e. the increase of pectin charge, gives a progressive increase in globule size. For DM0, the size increases from 110 to 185 up to 485 Å (the determination of $R_{comp}$ is less accurate when it approaches 500Å). For DM43, it also increases from 280 Å for pH 3 to 315 Å for pH 4 and 425 Å at pH 7. The globules of 400 Å and larger are less organized (we showed just above that the correlation peak is less narrow) but, surprisingly, the apparent inner globule composition decreases by only a few per cent.

On the contrary of pH, the effect of DM, which as we know also varies the charge, is not systematic. On the one hand, Table 2 shows that for pH 7, when increasing pectin linear charge density through DM, the globule size increases from 405Å (DM74) to 415 Å (DM43) and 485Å (DM0) like when increasing charge through the pH. On the other hand, for pH ~ 4 and pH ~3, the effect of DM is opposite. This results also in the fact that for the pair of samples corresponding to $[-]/[+]_{intro}$ ~3.5 (DM0 at pH 3.75 and DM43 at pH 7.2), $R_{comp}$ is much lower for DM0. We recall that in this case, the globules appear (from the SANS profile) to aggregate in larger structures instead of keeping on growing. This could find its origin from the fact that the DM0 chains are much shorter than the DM43 ones. It is likely that longer chains tend to form larger globules, because they cannot belong to a single small globule [28]. However, another pair of samples involving chains of the same, higher, molecular weight shows a similar effect! With equal $[-]/[+]_{intro}$ ~ 1.6 (DM43 at pH 4.1 and DM74 at pH 7.1), the two scattering curves are close (Fig. 4.a), but, looking at numbers, we find 405 Å for DM74, and 325 Å for DM43. It is possible that two effects are competing, so that obviously more data are necessary to a more systematic analysis.

Finally, the effect of the ionic strength I cannot be studied since at constant buffer ionic strength 25mM, the only change comes from the release of counterions of lysozyme and pectin: this is negligible except for $P_0$ at pH 7, I reaching 56mM. But in this latter case the strongest effect is most probably due to the huge increase in charge ratio ($[-]/[+]_{intro}$~ 10).

*5.2. Comparison with other proteins-polyelectrolyte systems*

The amount of knowledge available on the local structure of PSSNa/lysozyme mixtures [11, 15, 24, 28,] obtained by SANS owing to deuteriation labeling of PSS enables to discuss the main features of complexation by comparison with pectin/lysozyme systems because the protein is the same in both systems. There are strong analogies between the scattering described here on pectin/lysozyme and the scattering of PSSNa/lysozyme system, in the region of the phase diagram where the PSS chains are in a dilute regime after interacting with the proteins [36]. Macroscopically, their aspect is similar just after mixing in turbidity, and in viscosity, though this of course may cover a wide range of different structures. Anyway, the spatial organisation at the scale probed by SANS is similar. We obtain very similar globule scattering and protein stacking since we get the correlation peak around 0.2 Å$^{-1}$, a $q^{-4}$ law at intermediate q followed at lower q by a Guinier range for the globules. The sizes and the compactness can be obtained accurately in both systems. For the same lysozyme concentration (~ 5g/L), for a comparable charge ratio [-]/[+]$_{intro}$ (1.6) and similar ionic strength, one gets a much lower $R_{comp}$ for PSS/lysozyme (50Å [24]) than for DM43 and DM74. Hence much larger sizes are found for pectin at the same I. Concerning compactness, a value always higher than 0.25 is found for PSS/lysozyme whatever the conditions [24], hence higher than for pectin by at least more than a factor two (even 3 or more).

These differences must be due to the effect of the persistence length, in accordance with simulations [19, 20] which predict less complexation for more chain stiffness. The pectin has indeed a much larger intrinsic persistence length $L_0$ (80 Å) than PSS ($L_0$ = 12Å). In PSS/lysozyme systems, the persistence length of the chains, which can be as large as 100Å in pure solutions, could be reduced down to 17 Å when interacting with lysozyme, favouring dense globules [29, 33]. But this reduction is not possible for rigid pectin; this explains lower compacities. These results have to be put in parallel with the ones obtained on BSA-chitosan [34] and BSA-PDAMAC [35] coacervates. Very nicely, we obtain the same compacities in PEL-proteins systems for two different proteins, with different sizes and different charges, either when complexed with a rigid natural PEL (0.15 for lysozyme with pectin, 0.15 for BSA with chitosan) or when complexed with a flexible PEL (0.3 for lysozyme with PSS, 0.25 for BSA with PDAMAC). The stiffness of PEL thus seems to be a main parameter that controls the compactness of globules of PEL and proteins of opposite charges. It is especially important for applications because it governs the accessibility to the core of the globules. The effect of the stiffness can also play a role on the size of the globules. As noted above, the use of long chains can increase the size of the globule when a chain that belongs to several small globules bridges them to form a larger one. This effect could be enhanced with increasing persistence length and lead to higher globules size in pectin/lysozyme systems.

The semiflexible nature of pectin does not induce any elongated shape of the complexes, of the beads-on-a-string type. Thus we have neither chains decorated by proteins, nor parts of chains wrapping the lysozyme like in the artificial chromatin models predicted in case of partial aggregation for non stoechiometric charge ratio (x far from 1)[44]. The wrapped structure may be limited by the small protein radius (20 Å) compared to the persistence length (100 Å). Anyway, here we see only condensed globules, and never partial aggregation, even though the charge ratio is varied over a rather wide range (while being limited to constant chain and spheres concentration).

Another difference between the pectin/lysozyme system and the PSS/lysozyme systems lies in the low q region where we can compare the correlations between globules. They are attractive in both systems, but there can be two different ways. For PSS globular regime, the intermediate $q^{-4}$ scattering always crosses over to a $q^{-2.1}$ apparent fractal scattering law at low q [11, 31], attributed to the formation of dense aggregates of globules formed through a Reaction Limited Aggregation Process. For pectin, we observe for DM0 and 43, the most charged and the smallest globules, some resemblance with PSS. For larger pectin-lysozyme globules, we still observe attraction between globules (see Figure 6) but in the available q range the structure factor S(q) stays much closer to one. This could be due simply to the fact that the sizes $R_{comp}$ are much larger than for PSS-lysozyme (where they lie in between 75 Å and 175 Å). However, S(q) at low q in Figure 6 reaches a plateau, instead of increasing as does a power law function. The lower attraction could be related to the difference of surface charge densities of the globules in the two systems: the surface charge density for pectin being lower since the linear charge of pectin is weaker than the one of PSS. This would lead to softer attractions, which can eventually result in gas-liquid transitions of globules (coacervation [6]). The influence of electrostatics interactions on the structure of aggregates has also been studied by SAXS on whey protein isolate /gum arabic coacervates by Weinbreck et al [42]. Tuning the ionic strength instead of the aggregates charge density, they observed also that a screening of electrostatics by adding salt leads, through the increase of osmotic compressibility of the system, to a more heterogeneous and less structured coacervate.

A last interesting point of the comparison between the pectin/lysozyme system and the PSS/lysozyme concerns the hydrophobic effects. In the case of PSSNa mixed with lysozyme [28], it was found that at high concentration of PSSNa (for $[-]/[+]_{intro}$ > 5 to 10), lysozyme was unfolded. This unfolding was attributed to hydrophobic interaction between lysozyme and the phenyl ring of PSS. In the case of pectin studied here, the lysozyme keeps a perfectly globular conformation - the characteristic $q^{-4}$ decay of a compact protein at large q is always recovered, even for a DM of 74. Thus methyl is not aggressive like phenyl, in term of protein conformational change.

## *6. Summary and conclusion*

In summary SANS has allowed the characterization of the shape, internal organization and water content of lysozyme/pectin complexes in a part of the phase diagram. The complexes are made of globules with a radius of a few tens nanometers, except for very low charge ratios of the pectin (high degree of methylation DM and low pH) for which they have the structure of an inhomogeneous gel. Using absolute $cm^{-1}$ units for the cross-section, we obtain a volume fraction of protein plus pectin in the globules of around 10%, ± 3% depending on the different parameters probed in the study (DM, PH, molecular weight), i.e. a water content of about 90%. On small scales, proteins are in close contact, and such stacking inside the globules depends mostly on the DM, i.e. the distance between charges monitors the compactness of the chain–protein arrangement inside the globules. The protein stacking varies in a way opposite to the one we would have expected for the case of hydrophobic interactions: we conclude that hydrophobicity plays a minor role in the complexation. On large scales, the sizes of the globules vary between 10 nm, for the lowest pectin degree of charge and 50 nm for the highest. The linear chain density seems to be the controlling parameter, when it is controlled by the pH, but the variation is contradictory when the DM is varied. The comparison is complicated by the fact that at low DM (lower charge) the globules remain small and aggregate into larger structures. This may be due to the lower molecular weight of the DM0chain: keeping all other physico-chemical parameters constant, short chains tend to form smaller globules).

We can also conclude that we do not observe any beads-on-a-string structure as predicted for partial aggregation [44].

Comparisons with others PEL/proteins complexes show that we can obtain very general trends for the influence of chain stiffness on the compactness of complexes, an important feature for potential applications as it governs the accessibility of the inner of the globules: flexible chains lead to dense globules (water content is ~ 70% as observed in PSS/lysozyme system [24] and BSA/PDAMAC [34]), while rigid chains lead to globules with a high content of water (~ 90% as observed here on PSS/pectin system and 85% in BSA/chitosan system [35]). This is in accordance with simulations.

The type of SANS analysis we develop here could be extended to other complexes between other proteins and other polyelectrolytes, in particular other polysaccharides.

**Supporting Information**: I Charge of components of the mixture as a function of pH. II Neutron scattering length densities of the components. III Volume of protein obtained from SANS experiments. IV Test of contrast matching of pectin or lysozyme within globules. V. Comparison at given DM : enlargement of Figure 3 and test of reproducibility using two pairs of samples prepared at same pH, measured and treated separately (at different dates).

This material is available free of charge via the Internet on: http://pubs.acs.org.

**Appendix:** Compactness of primary globules extracted from the Porod regime

The two ingredients of the trick is based on the plot $I(q).q^4/2\pi = f(q)$ of the scattering curves (Figure 5). Once having determined, from the X axis, the **value of $R_{comp}$**, it is possible to use the absolute values along the Y coordinate, $I(q).q^4/2\pi$ to obtain **the contrast of the globules**: $\Delta\rho_{comp}^2$ can be deduced if we assume that the globules are spheres. The Porod radius $R_p$ of globules is identified to its real radius $R_{comp}$. We define $\Phi_{comp}$ as the volume fraction occupied by the globules in the sample, we get:

$$I(q).q^4 / 2\pi = \Delta\rho_{comp}^2 \, S/V \qquad (A.1)$$

As $S/V = (1-\phi_{comp}) \cdot \phi_{comp} \, S_{comp}/V_{comp}$ and $S_{comp}/V_{comp} = 3/R_p$, we obtain:

$$\Delta\rho_{comp}^2 = I q^4 \, R_{comp} / 6\pi \, ((1-\phi_{comp}) \, \phi_{comp}) \qquad (A.2)$$

We see that the product $(R_{comp} \cdot I(q) \cdot q^4)/2\pi$ is constant for mixtures showing well defined globules. This is a simple way of checking that $\Delta\rho_{comp}^2 (1-\Phi_{comp})\Phi_{comp}$ is constant. Let us discuss this in more details. The globules are composed of water, lysozyme and pectin. The volume fraction of water within the globules is $\Phi_{D20\_comp}$. If we assume that all pectins and lysozyme are inside the complexes, in accordance with UV titrations on lysozyme, and hence in the globules, we get:

$$\Phi_{comp} = \Phi_{lyso} + \Phi_{pectin} + \Phi_{D20\_comp} \quad \text{(A.3)}$$

We define the volume fractions of lysozyme and pectin within the globules by:

$$\Phi_{lyso\_comp} = \Phi_{lyso}/\Phi_{comp}, \quad \Phi_{pectin\_comp} = \Phi_{pectin}/\Phi_{comp}.$$

The neutronic contrast between the globules and the solvent writes:

$$\Delta\rho_{comp}^2 = (\Phi_{lyso\_comp}(\rho_{D20} - \rho_{lyso}) + \Phi_{pectin\_comp}(\rho_{D20} - \rho_{pectin}))^2 \quad \text{(A.4)}$$

$$= (1/\Phi_{comp})^2(\Phi_{lyso}(\rho_{lyso} - \rho_{D20}) + \Phi_{pectin}(\rho_{pectin} - \rho_{D20}))^2 \quad \text{(A.5)}$$

and from the $q^4 I(q)$ values we extract:

$$\Delta\rho_{comp}^2 = (Iq^4/2\pi) \cdot R_{comp}/3(1-\Phi_{comp})\Phi_{comp} \quad \text{(A.6)}$$

So from (A.5), we obtain

$$(1-\Phi_{comp})\Phi_{comp} \cdot (1/\Phi_{comp})^2(\Phi_{lyso}(\rho_{lyso} - \rho_{D20}) + \Phi_{pectin}(\rho_{pectin} - \rho_{D20}))^2$$

$$= (Iq^4/6\pi) \cdot R_{comp}$$

Or in other terms:

$$(1-\Phi_{comp})/\Phi_{comp} = (Iq^4 \cdot R_{comp})/6\pi(\Phi_{lyso}(\rho_{lyso} - \rho_{D20}) + \Phi_{pectin}(\rho_{pectin} - \rho_{D20}))^2$$

Hence $\Phi_{comp}$ can be written:

$$\Phi_{comp} = \cfrac{1}{1 + \cfrac{\frac{Iq^4 R_{comp}}{6\pi}}{(\Phi_{lyso}(\rho_{D2O} - \rho_{lyso}) + \Phi_{pectin}(\rho_{D2O} - \rho_{pectin}))^2}} \quad \text{(A.7)}$$

If in addition we assume that all organic species are localized inside the globules, we then we can also get the volume fraction of protein plus pectin within the globules:

$$\Phi_{inner} = \Phi_{lyso\_comp} + \Phi_{pectin\_comp} \quad (= (\Phi_{lyso} + \Phi_{pectin})/\Phi_{comp}). \quad \text{(A.8)}$$

**Figure captions.**

Figures are in a separate file.

Figure 1: SANS spectra of individual components of the mixture: ■ lysozyme (10g/L); ▲ pectin DM74 (10g/L).

Figure 2: SANS spectra for lysozyme-pectin complexes for each of the three pectins (DM0, DM43 and DM74) as a function of pH: (a) pH ~ 7, (b) pH ~ 4, (c) pH ~ 3. Each set of spectra is compared to the spectrum of free lysozyme at the same concentration as in the mixture (5g/L).

[Figure 3]: SANS spectra for lysozyme-pectin complexes for the three pH values (pH ~3, pH ~4 and pH ~7) as function of DM, at large q : (a) DM 0, (b) DM 43, (c) DM 74. and at low q (d) for DM 0. Each set of spectra is compared to the spectrum of free lysozyme at the same concentration as in the mixture (5g/L).

Figure 4: Comparison of SANS spectra for lysozyme-pectin complexes at constant charge ratio (a) $[-]/[+]_{intro}$ ~ 3.5 and (b) $[-]/[+]_{intro}$ ~ 1.5.

Figure 5: Two examples of $I(q)q^4/2\pi = f(q)$ representations at constant DM and constant pH (a) DM 0 at all pH, lower q range; (b) all DMs at pH 7, full q range.

Figure 6: Structure factor S(q) for mixtures of lysozyme and pectin at very low q at pH 7 for degree of methylation 74 and 43, at which the globules are the best defined.

Figure 7: Tentative sketch of globular pectin-lysozyme complexes and their scattering in log-log plot (the $q^{-Dapp}$ part at low q is seen only in DM0 samples).

**References.**

[1] Bungenberg de Jong, H.G. *La coacervation complexe et son importance en biologie*, Fauré - Fremiet, E. Ed; Hermann et Cie: Paris, 1936.

[2] Bungenberg de Jong, H.G. *Crystallisation – Coacervation – Flocculation*, In: Colloid Science, Kruyt, H.R., Ed.; Elsevier: Amsterdam, 1949, vol 2, p 232.

[3] Morawetz, H.; Hughes, W. L. *J. Phys. Chem.* **1951**, *56*, 64-69.

[4] Tribet, C. *Complexation between Amphiphilic Polyelectrolytes and Proteins: from Necklaces to Gels*, Surfactant Science Series "Physical chemistry of polyelectrolytes"; Radeva, T., Ed.; M. Dekker: New York, 1999; Chapter 19, pp 687-741.

[5] Doublier, J.L.; Garnier, C. ; Renard, D.; Sanchez, C. *Curr. Opin. Colloid Interface Sci.* **2000**, *5*, 202-214.

[6] de Kruif, C.G.; Weinbreck, F.; deVries, R. *Curr. Opin. Colloid Interface Sci.* **2004,** 9, 340-349.

[7] Cooper, C.L.; Dubin, P.L.; Kayitmazer, A.B.; Turksen, S. *Curr. Opin. Colloid Interface Sci*. **2005**, *10*, 52-78.

[8] Turgeon, S.L.; Schmitt, C.; Sanchez, C. *Curr. Opin. Colloid Interface Sci*. **2007**, *12*, 166-178.

[9] Jourdain, L.; Leser, Martin E.; Schmitt, C.; Michel, M.; Dickinson, E. *Food Hydrocolloids* **2008**, *22*, 647–659.

[10] Ball, V.; Winterhalter, M.; Schwinte, P.; Lavalle, Ph.; Voegel J.-C.; Schaaf, P. *J. Phys. Chem. B* **2002**, 106, 2357-2364.

[11] Gummel, J.; Boué, F.; Demé, B.; Cousin, F. *J. Phys. Chem. B* **2006**, *110*, 24837-24846.

[12] Park, J.M.; Muhoberac, B.B.; Dubin, P.L. ; Xia, J. *Macromolecules* **1992**, 25, 290-295.

[13] Tsuboi, A.; Izumi, T.; Hirata, M.; Xia, J.; Dubin, P. J.; Kokufuta, E. *Langmuir* **1996**, *12*, 6295-6303.

[14] Schmitt, C.; Palma da Silva, T.; Bovay, C., Rami-Shojaei, S., Frossard, P., Kolodziejczyk, E., Leser M.E. *Langmuir* **2005**, *21*, 7786-7795

[15] Gummel, J.; Cousin, F.; Boué, F. *J. Am. Chem. Soc.* **2007,** *129* (18), 5806-5807.

[16] Turgeon, S.L.; Beaulieu, M.; Schmitt, C ; Sanchez, C. *Curr. Opin. Coll. Interface Sci*. **2003,** *8*, 401-414.

[17] de Kruif, C.G.; Tuinier, R. *Food Hydrocolloids,* **2001,** *15*, 555-563.

[18] Girard, M.; Turgeon, S.L. ; Gauthier, S. F. *J. Agric. Food Chem.* **2003,** *51*, 4450-4455.

[19] Stoll, S.; Chodanowski P. *Macromolecules* **2002**, *35*, 9556-9562.

[20] Ulrich, S.; Laguecir, A.; Stoll S. *Macromolecules* **2005,** *38*, 8939-8949.

[21] Carlsson F., Malmsten M., Linse P. *J. Am. Chem. Soc.* **2003,** *125*, 3140-3149.

[22] Ulrich, S.; Seijo M.; Laguecir, A.; Stoll S. *J. Phys. Chem. B* **2006,** *110*, 20954-20964.

[23] Kayitmazer, A. B.; Shaw, D.; Dubin, P. L. *Macromolecules*, **2005**, 38(12), 5198-5204.

[24] Gummel, J.; Cousin, F.; Clemens, D.; Boué, F. *Soft Matter* **2008**, *4*, 1653–1664.

[25] Seyrek, E.; Dubin, P. L.; Tribet, C.; Gamble, E. A. *Biomacromolecules*, **2003**, *4(2)*, 273-282.

[26] Borrega, R.; Tribet, C.; Audebert, R. *Macromolecules* **1999,** *32*, 7798-7806.

[27] F. Petit, R. Audebert, I. Iliopoulos, *Colloid Poly. Sci.* **1995,** *273*, 777-781.

[28] Cousin, F.; Gummel, J.; Ung, D.; Boué, F. *Langmuir* **2005,** *21*, 9675-9688.

[29] Sanchez, C.; Mekhloufi, G.; Renard, D. *J. Coll. Interface Sci.* **2006,** 299, 867-873.

[30] Girard, M.; Sanchez, C; Laneuville; S.I.; Turgeon; S.; Gauthier, S. *Colloids and Surfaces B*, **2004**, *35*, 15-22.

[31] Gummel, J.; Cousin, F.; Verbavatz, J-M.; Boué, F. *J. Phys. Chem. B* **2007,** *111*, 8540-8546.

[32] Wang, X.; Li, Y.; Wang, Y.-W.; Lal, J.; Huang, Q. *J. Phys. Chem. B* **2007,** *111*, 515-520.

[33] Santinath Singh, S.; Aswal V.K.; Bohidar H.B.; *Int. J. Biological Macromolecules* **2007,** *41*, 301-307.

[34] Kayitmazer, A.B.; Sabina P. Strand, S. P. ; Tribet, C.; Jaeger, W., Dubin, P.L., *Biomacromolecules*, **2007**, *8*, 3568-3577.

[35] Bohidar, H.; Dubin, P. L.; Majhi, P. R.; Tribet, C.; Jaeger, W. *Biomacromolecules*, **2005**, *6*, 1573-1585.

[36] Gummel, J.; Cousin, F.; Boué, F. *Macromolecules* **2008,** *41*, 2898-2907.

[37] Protein data bank; http://www.rcsb.org/pdb/home/home.do

[38] Cros, S.; Garnier, C.; Axelos, M.A.V.; Imberty, A.; Perez, S. *Biopolymers* **1995,** *39* , 339-352.

[39] Ralet, M.-C.; Crépeau, M.-J.; Lefèbvre, J.; Mouille, G.; Höfte, H.; Thibault, J.-F. *Biomacromolecules*, **2008**, *9(5)*, 1454-1460.

[40] Schmidt, I. *Structures et propriétés tensioactives des assemblages complexes protéines basiques / pectines*, Ph. D, Université Nantes, 2004.

[41] Cotton, J. P. *J. Phys. IV France* **1999,** *9*, 21-49.

[42] Weinbreck, F. ; Tromp, R.H. ; de Kruif C. G. *Biomacromolecules*, **2004**, *5*, 1437-1445.

[43] Rui Zhang, B.I. Shklovskii, *Physica A*, **2005**, *352*, 216-238.

[44] Rui Zhang, B.I. Shklovskii, *Phys. Rev. E* **2004,** *69*, 021909.

Figure 1

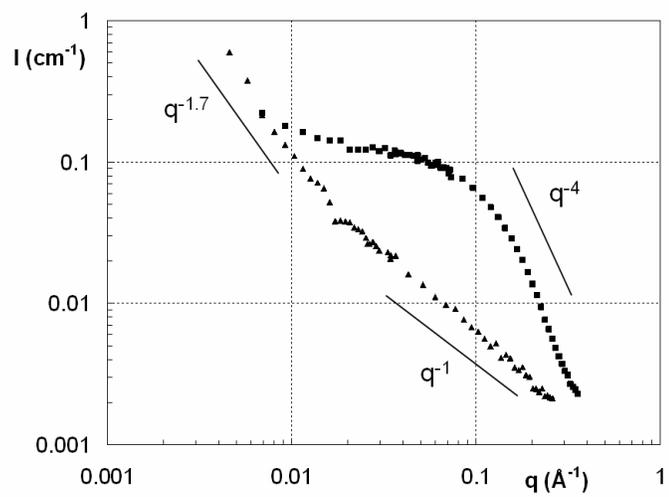

**Figure 1**

Figure 1: SANS spectra of individual components of the mixture: ■ lysozyme (10g/L); ▲ pectin DM74 (10g/L).

Figure 2a

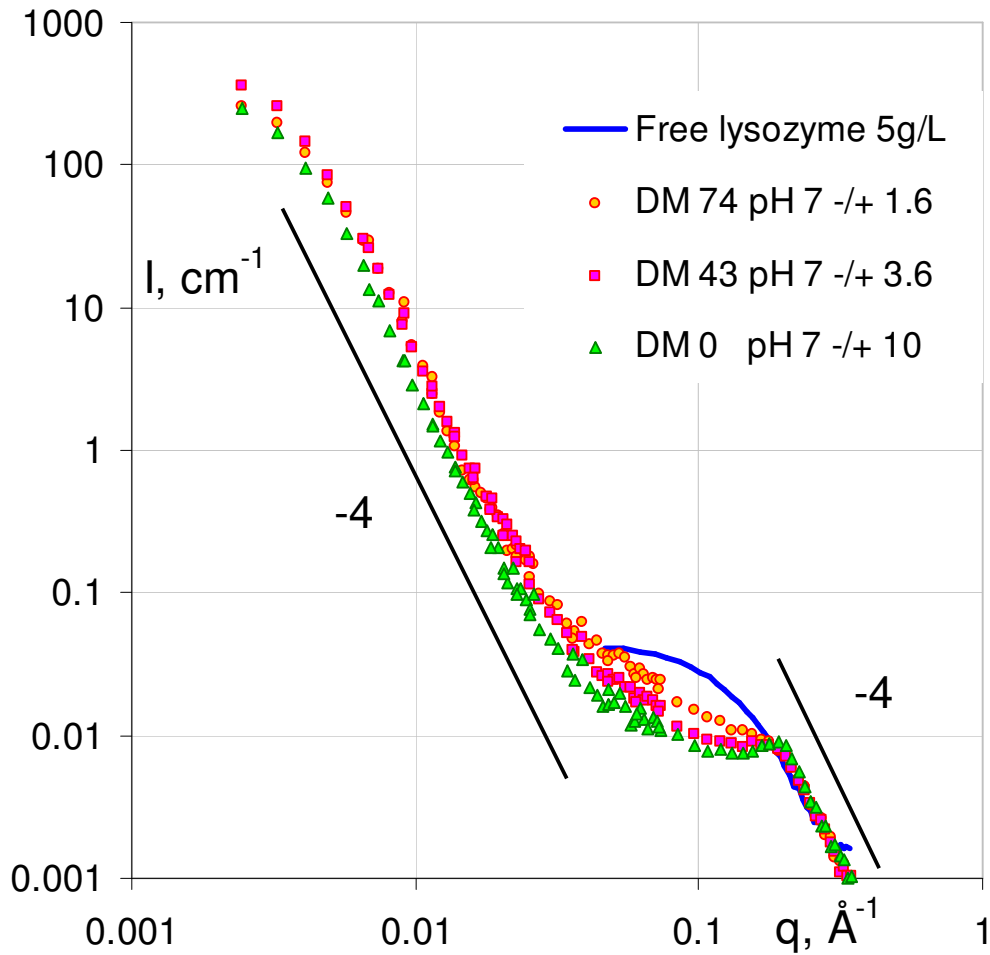

**Figure 2a**

Figure 2: SANS spectra for lysozyme-pectin complexes for each of the three pectins (DM0, DM43 and DM74) as a function of pH: (a) pH ~ 7, (b) pH ~ 4, (c) pH ~ 3. Each set of spectra is compared to the spectrum of free lysozyme at the same concentration as in the mixture (5g/L).

Figure 2b

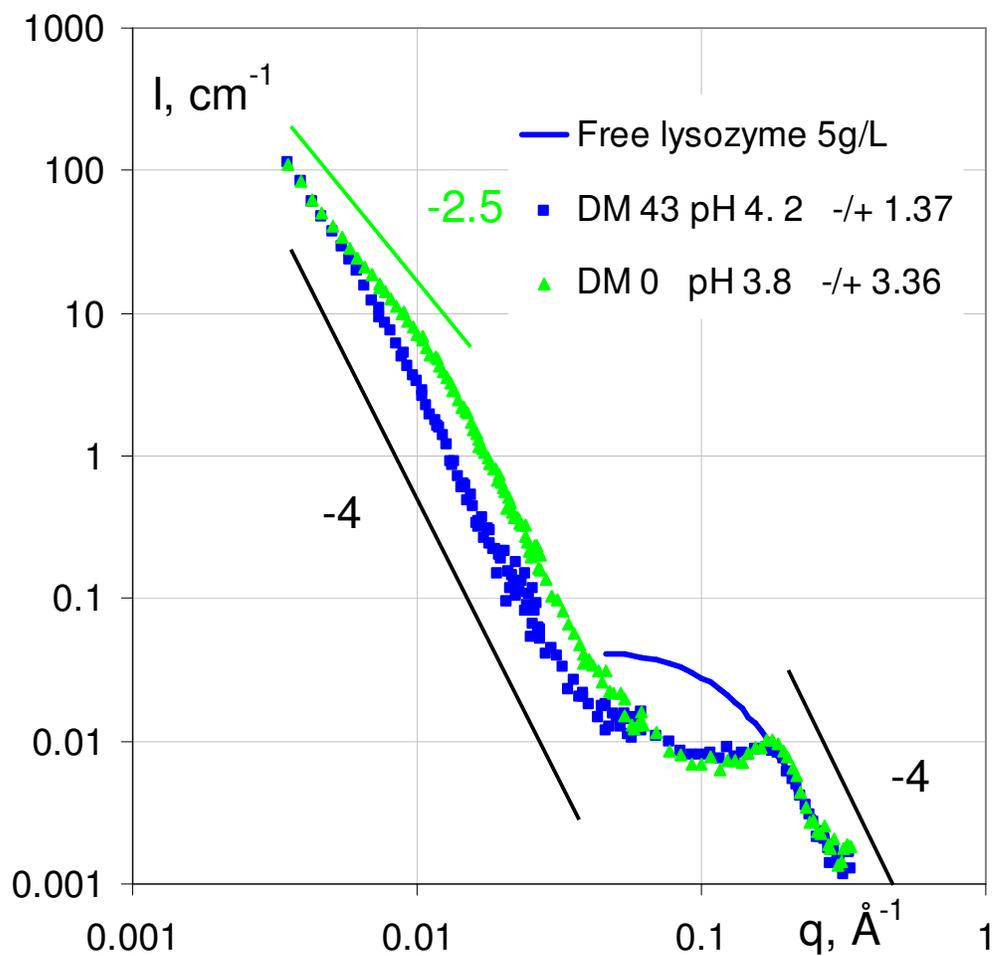

**Figure 2b**

Figure 2: SANS spectra for lysozyme-pectin complexes for each of the three pectins (DM0, DM43 and DM74) as a function of pH: (a) pH ~ 7, (b) pH ~ 4, (c) pH ~ 3. Each set of spectra is compared to the spectrum of free lysozyme at the same concentration as in the mixture (5g/L).

Figure 2c

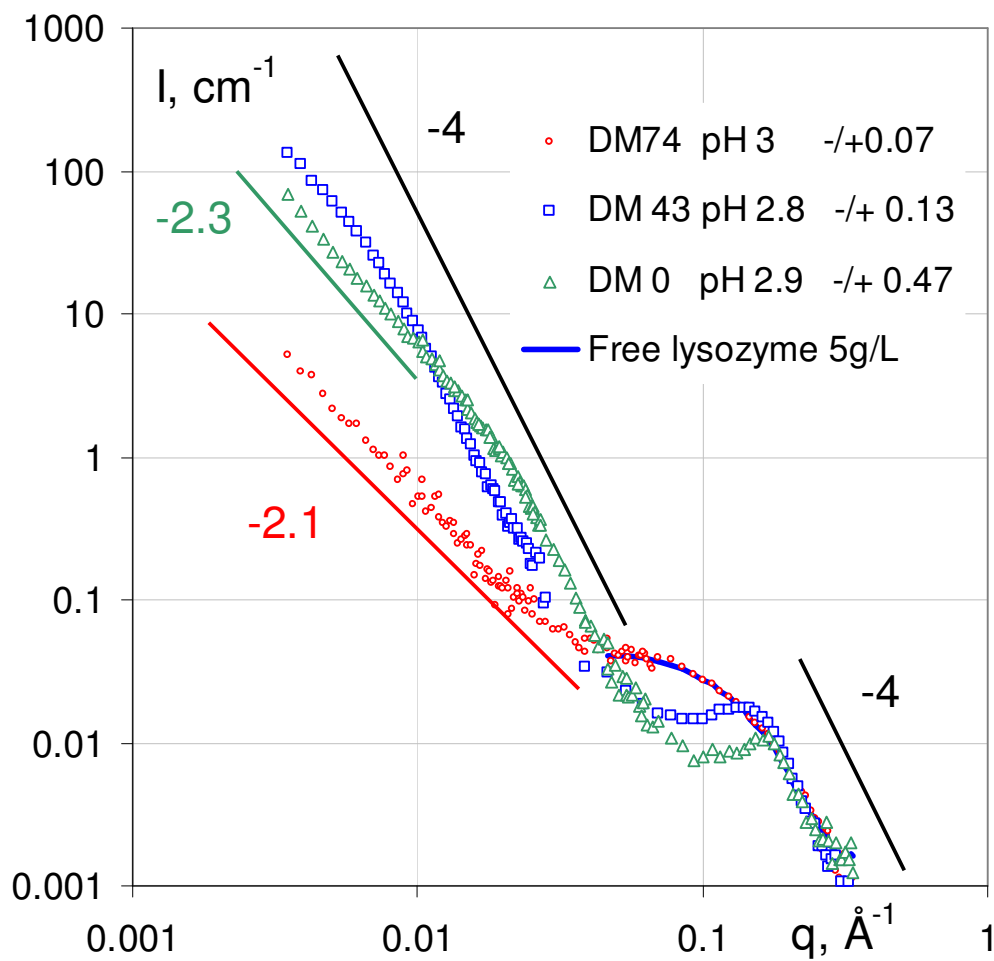

**Figure 2c**

Figure 2: SANS spectra for lysozyme-pectin complexes for each of the three pectins (DM0, DM43 and DM74) as a function of pH: (a) pH ~ 7, (b) pH ~ 4, (c) pH ~ 3. Each set of spectra is compared to the spectrum of free lysozyme at the same concentration as in the mixture (5g/L).

Figure 3a

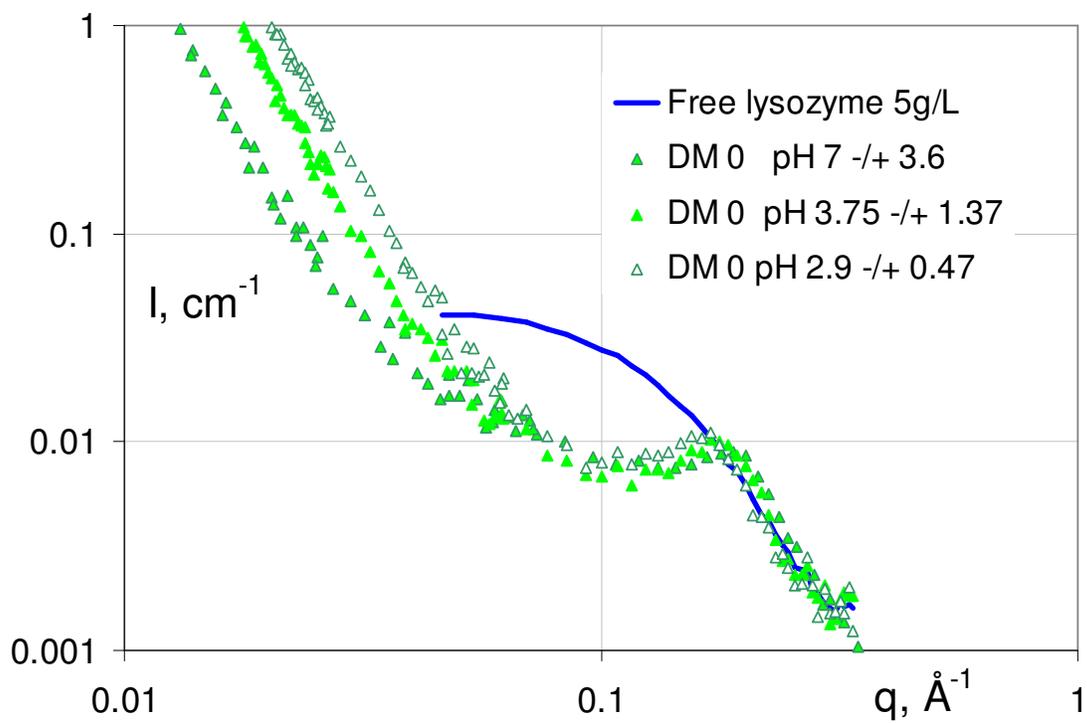

**Figure 3a**

Figure 3: SANS spectra for lysozyme-pectin complexes for the three pH values (pH ~3, pH ~4 and pH ~7) as function of DM at low q: (a) DM 0, (b) DM 43, (c) DM 74, and at low q (d) for DM0. Each set of spectra is compared to the spectrum of free lysozyme at the same concentration as in the mixture (5g/L).

Figure 3b

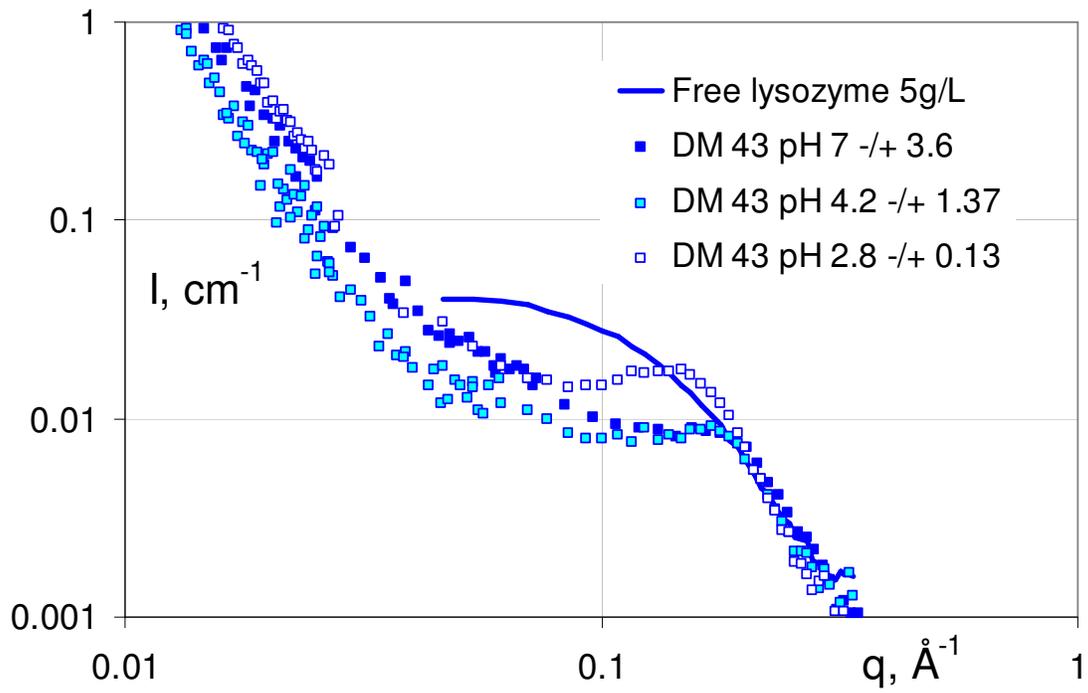

**Figure 3b**

Figure 3: SANS spectra for lysozyme-pectin complexes for the three pH values (pH ~3, pH ~4 and pH ~7) as function of DM at low q: (a) DM 0, (b) DM 43, (c) DM 74, and at low q (d) for DM0. Each set of spectra is compared to the spectrum of free lysozyme at the same concentration as in the mixture (5g/L).

Figure 3c

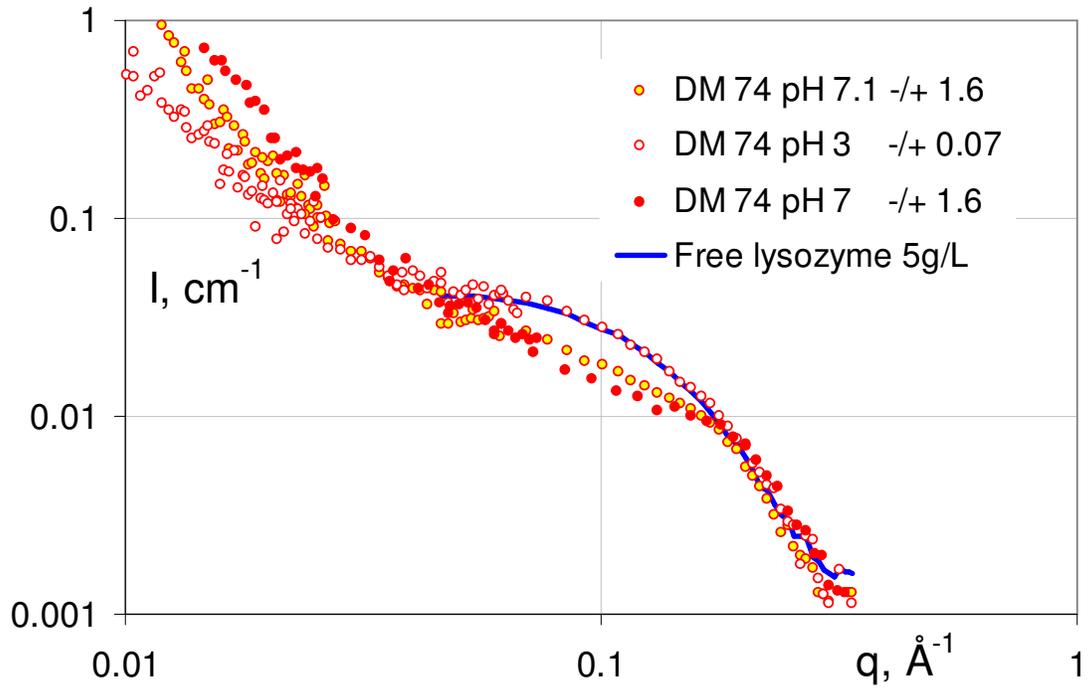

**Figure 3c**

Figure 3: SANS spectra for lysozyme-pectin complexes for the three pH values (pH ~3, pH ~4 and pH ~7) as function of DM at low q: (a) DM 0, (b) DM 43, (c) DM 74, and at low q (d) for DM0. Each set of spectra is compared to the spectrum of free lysozyme at the same concentration as in the mixture (5g/L).

Figure 3.d

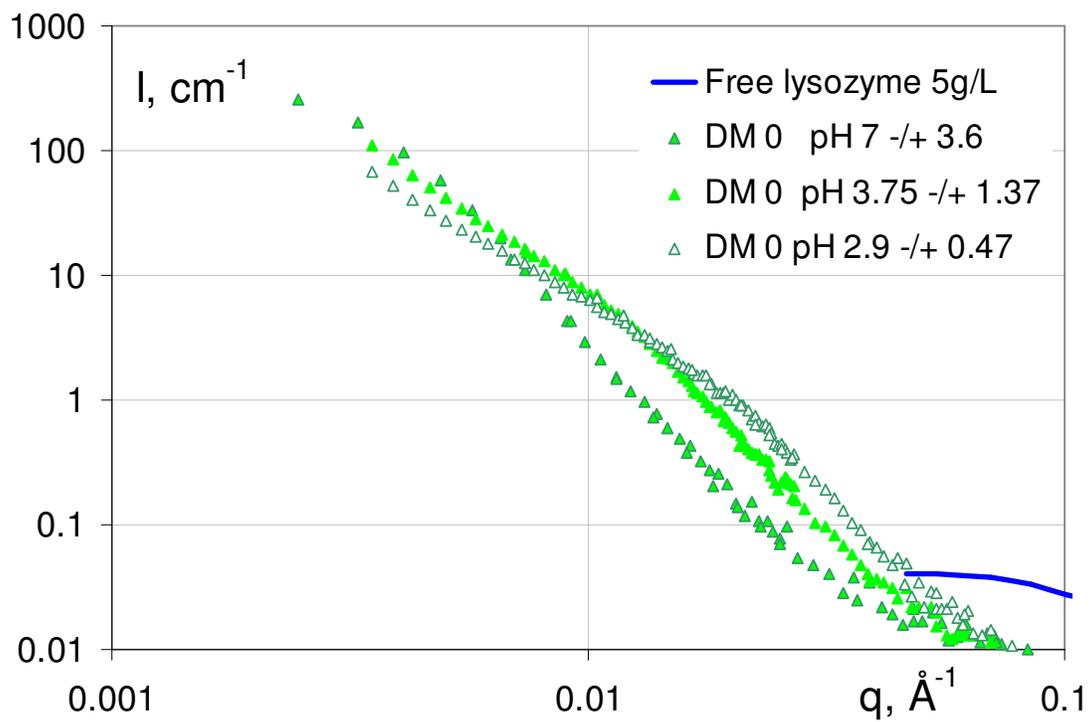

Figure 3: SANS spectra for lysozyme-pectin complexes for the three pH values (pH ~3, pH ~4 and pH ~7) as function of DM at low q: (a) DM 0, (b) DM 43, (c) DM 74, and at low q (d) for DM0. Each set of spectra is compared to the spectrum of free lysozyme at the same concentration as in the mixture (5g/L).

Figure 4a

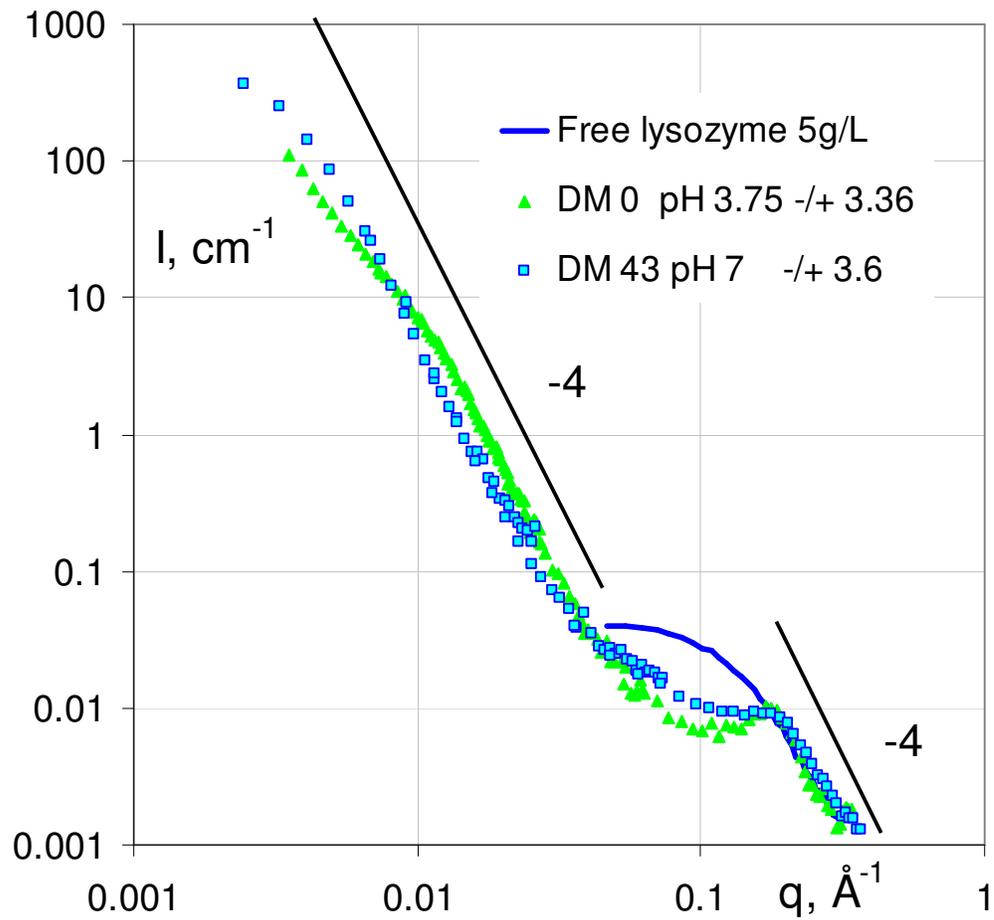

**Figure 4a**

Figure 4b

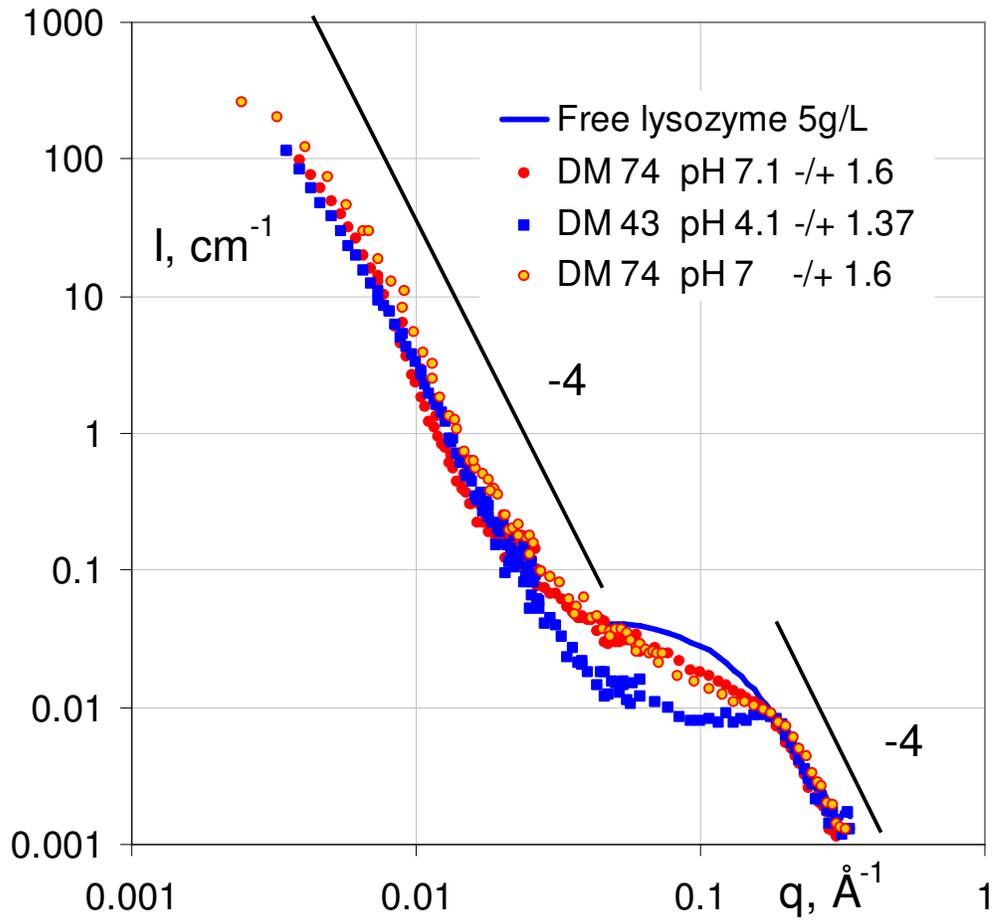

**Figure 4b**

Figure 5a

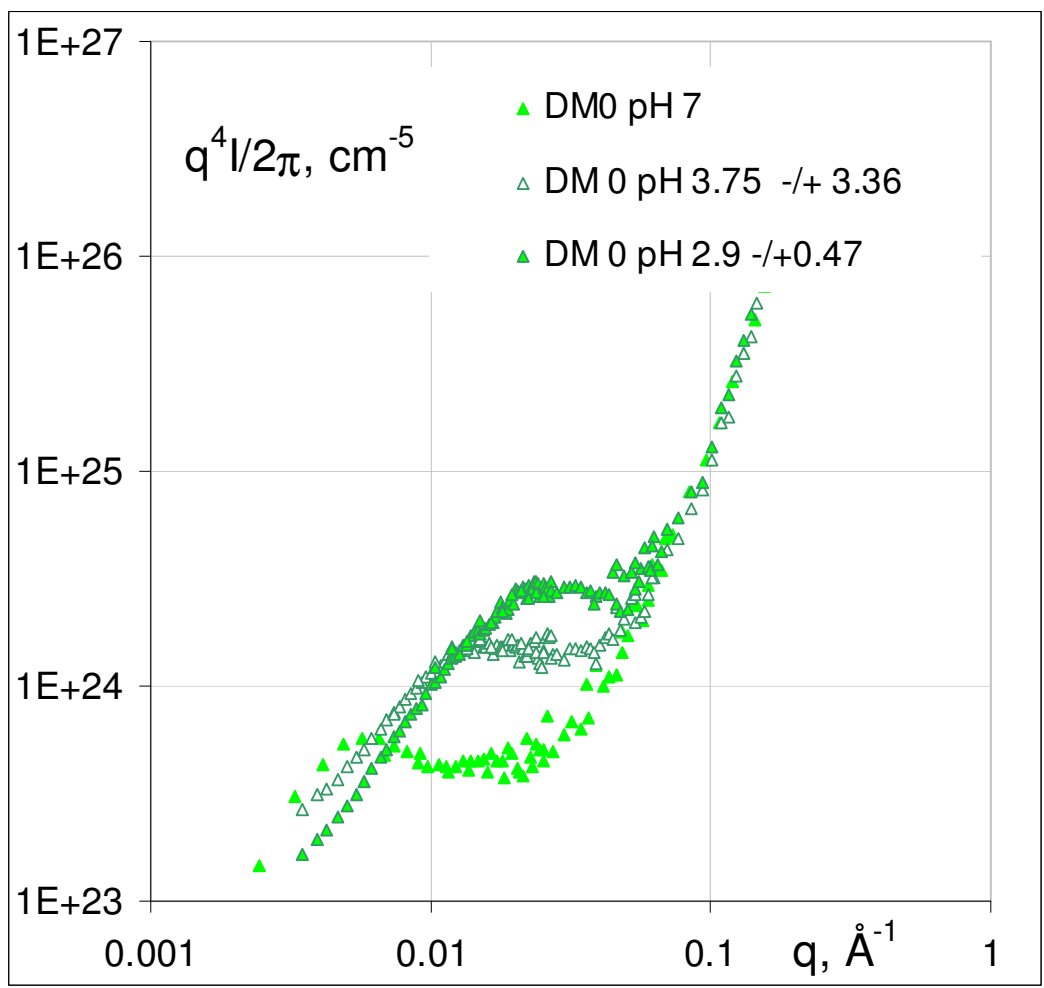

**Figure 5a**

Figure 5: Two examples of $I(q)q^4/2\pi = f(q)$ representations at constant DM and constant pH (a) DM 0 at all pH, lower q range; (b) all DMs at pH 7, full q range.

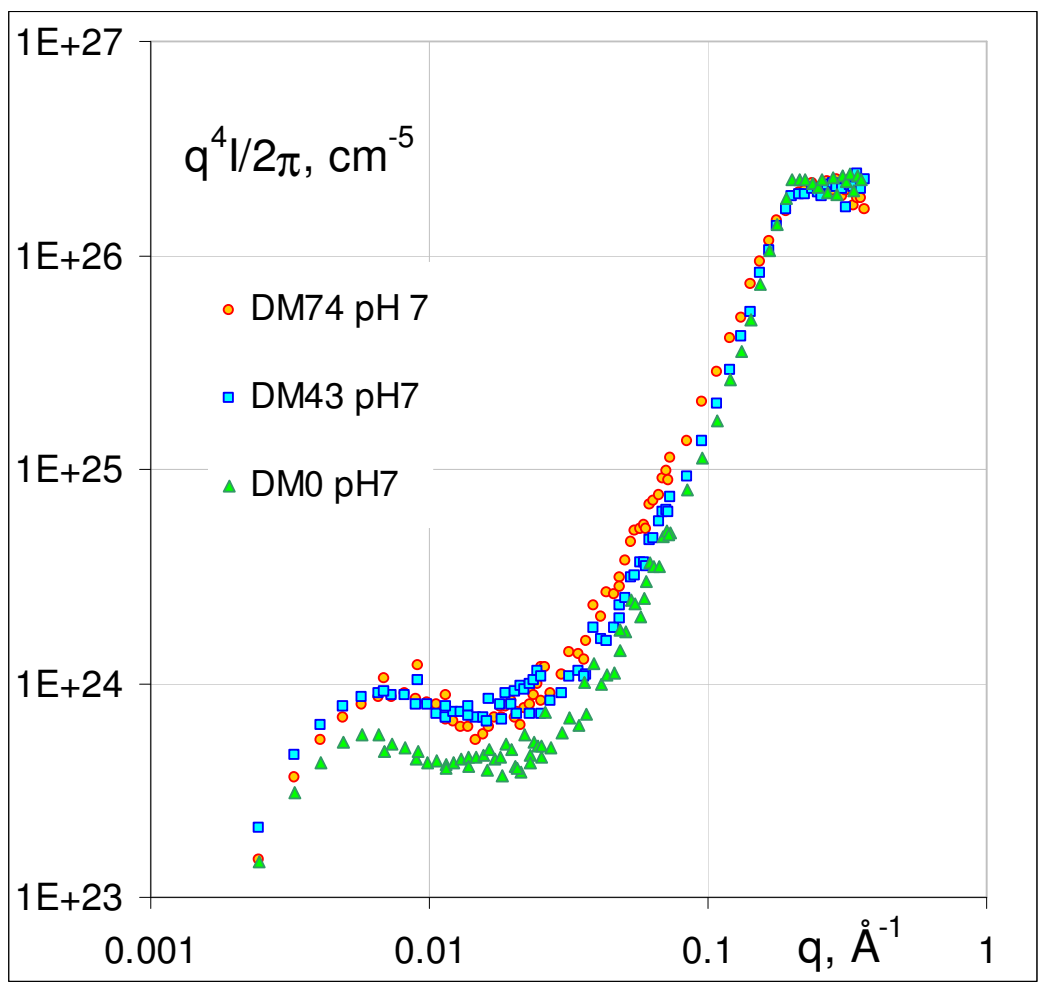

**Figure 5b**

Figure 5: Two examples of $I(q)q^4/2\pi = f(q)$ representations at constant DM and constant pH (a) DM 0 at all pH, lower q range; (b) all DMs at pH 7, full q range.

Figure 6

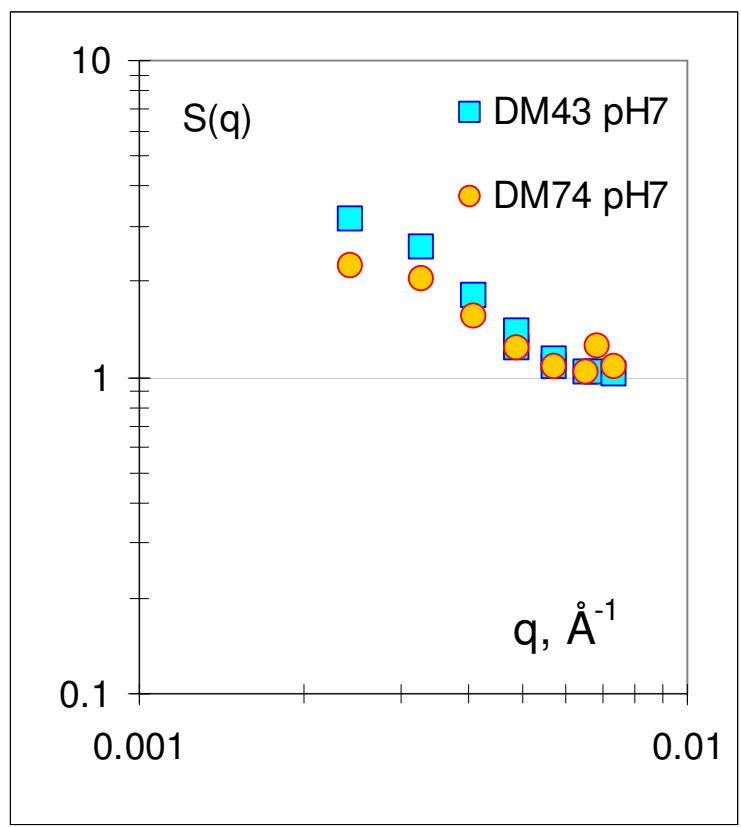

**Figure 6**

Figure 6: Structure factor S(q) for mixtures of lysozyme and pectin at very low q at pH 7 for degree of methylation 74 and 43, at which the globules are the best defined.

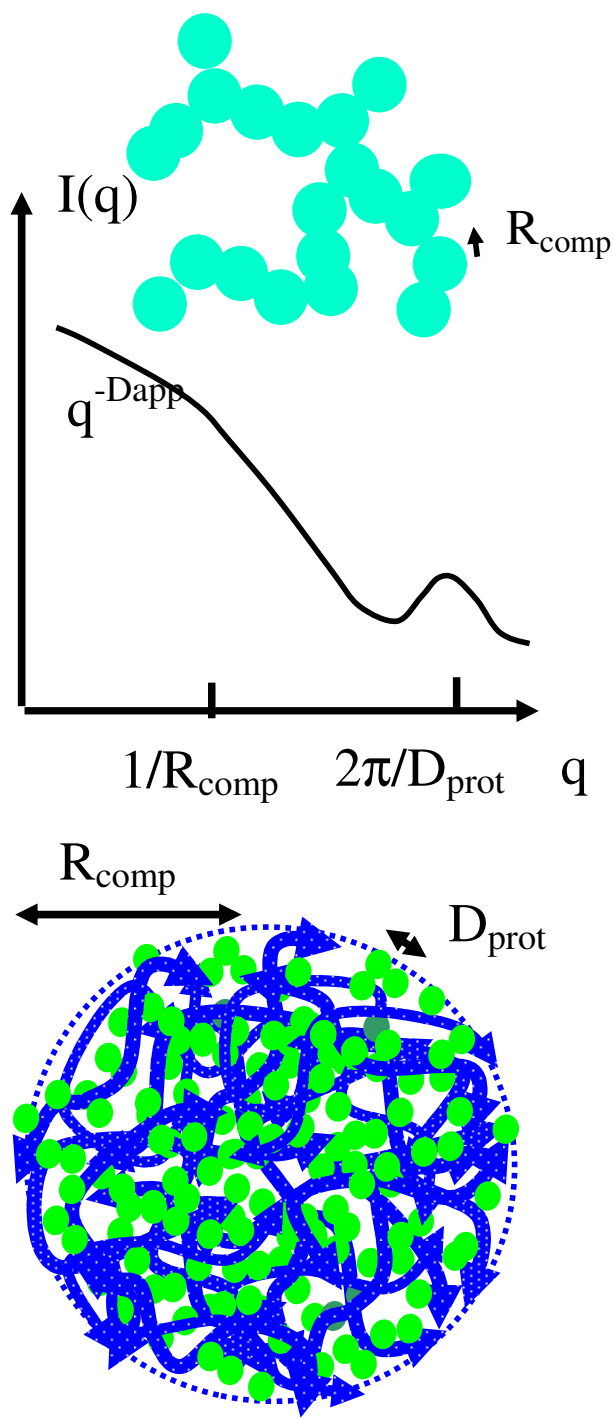

Figure 7 : Tentative sketch of globular pectin-lysozyme complexes and their scattering in log-log plot (the $q^{-D_{app}}$ part at low q from ramified aggregates of spheres is seen only in DM0 samples).